%% file: sample-journal.tex
 \documentclass[acmtog, authorversion]{acmart}

\usepackage{booktabs} 

\usepackage[ruled]{algorithm2e} 
\usepackage{amsmath,bm}

\usepackage{cuted}
\usepackage{capt-of}
\usepackage{graphicx}
\usepackage{algorithmicx}
\usepackage{textcomp}

\begin{document}

\title{A Parallel Feature-preserving Mesh Variable Offsetting Approach Based On Dynamic Programming}

 \titlenote{Corresponding author: gxu@hdu.edu.cn (Gang XU)}

\author{Hongyi Cao, Gang Xu *, Renshu Gu, Jinlan Xu}
\affiliation{%
  \institution{Hangzhou Dianzi University}
  \streetaddress{2\# Street 1158}
  \city{Hangzhou}
  \state{Zhejiang}
  \postcode{310018}
  \country{China}
}

\author{Xiaoyu Zhang}
\affiliation{%
  \institution{Beijing Institute of Spacecraft System Engineering}
  \city{Beijing}
  \country{China}
}
\author{Timon Rabczuk}
\affiliation{%
  \institution{Institute of Structural Mechanics}
  \city{Bauhaus-Universität Weimar}
  \country{Germany}
}

\begin{abstract}

Mesh offsetting plays an important role in discrete geometric processing. In this paper, we propose a parallel feature-preserving mesh offsetting framework with variable distance. Different from the traditional method based on distance and normal vector, a new calculation of offset position is proposed by using dynamic programming and quadratic programming, and the sharp feature can be preserved after offsetting. Instead of distance implicit field, a spatial coverage region represented by polyhedral for computing offsets is proposed. Our method can generate an offsetting model with smaller mesh size, and also can achieve high quality without gaps, holes, and self-intersections. Moreover, several acceleration techniques are proposed for the efficient mesh offsetting, such as the parallel computing with grid, AABB tree and rays computing. In order to show the efficiency and robustness of the proposed framework, we have tested our method on the quadmesh dataset, which is available at [https://www.quadmesh.cloud].
The source code of the proposed algorithm is available on GitHub at [https://github.com/iGame-Lab/PFPOffset].

\end{abstract}

%
%

%
%


\begin{teaserfigure}
\centering
\includegraphics[width=7.1 in]{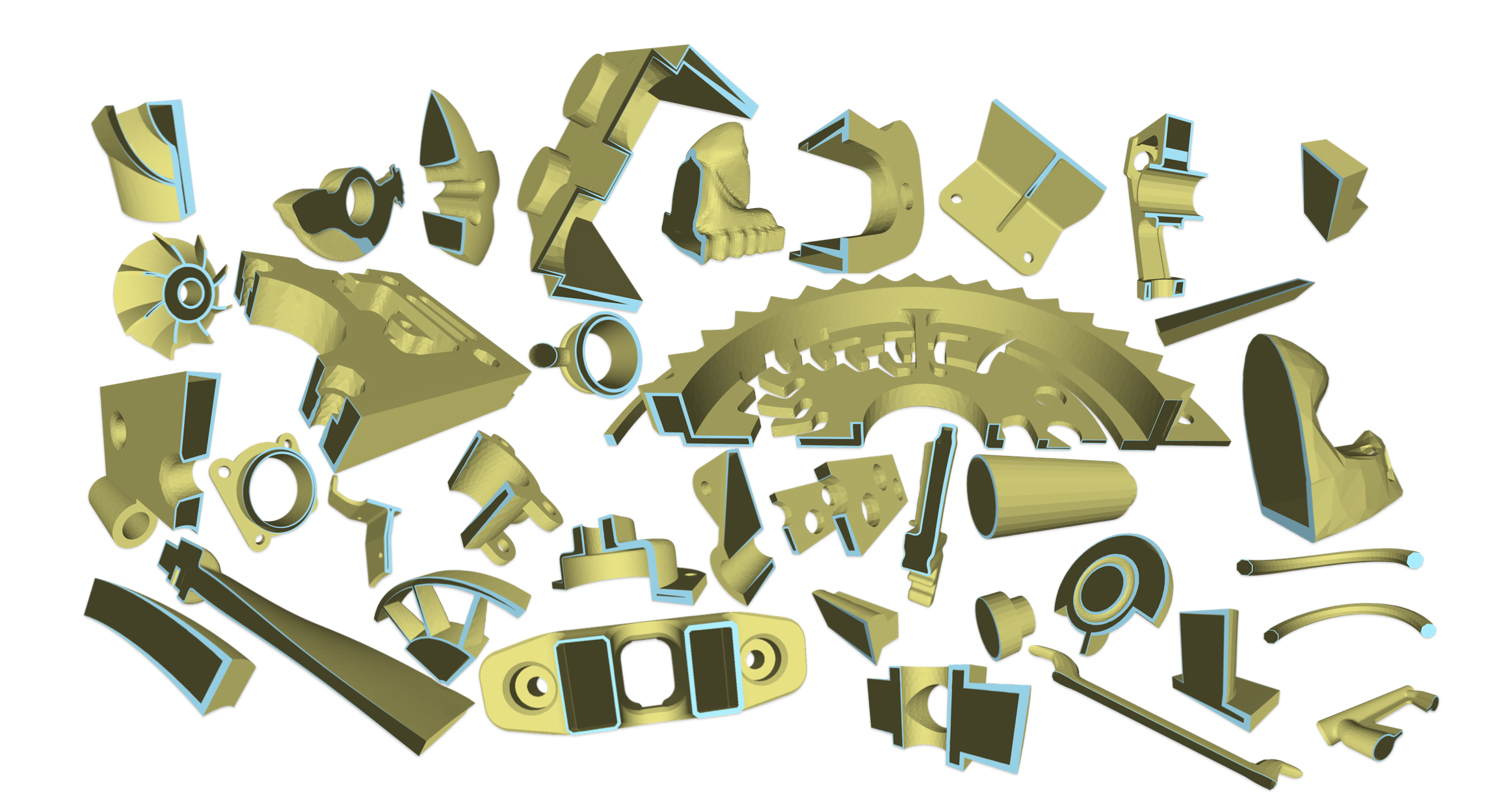}
\label{fig:teaser}
\end{teaserfigure}

\maketitle

\input{introduction.tex}

\input{RelatedWork.tex}

\input{Method.tex}
\input{ExperimentalResults.tex}

\input{results.tex}
\input{conclusion.tex}

\end{document}

%% file: introduction.tex
\section{Introduction}

Mesh offsetting is an important research topic in the field of digital geometric processing and has many applications, including but not limited to computer-aided design (CAD), collision detection, path planning, boundary layer mesh generation, and architectural design \cite{zint2023feature}. Surface mesh offsetting, for instance, offers a versatile tool for post-production adjustments to meshes, allowing sections to be enlarged or reduced based on user requirements or outcomes from physical simulations. Among them, offsetting with variable distance has many applications in CAD models of aviation devices. Further, this technique can morph a triangular mesh into a hollow object. The original triangular mesh can be viewed as the outer surface. By offsetting it inwardly, an inner surface emerges.
Together, these surfaces depict a solid entity with thickness, commonly termed a "shell-like solid". This structure is notably leveraged in additive manufacturing \cite{xie2017support}.

\begin{figure}[htb]
 \begin{minipage}[c]{0.5\textwidth}
    \centering
    \includegraphics[width=2 in]{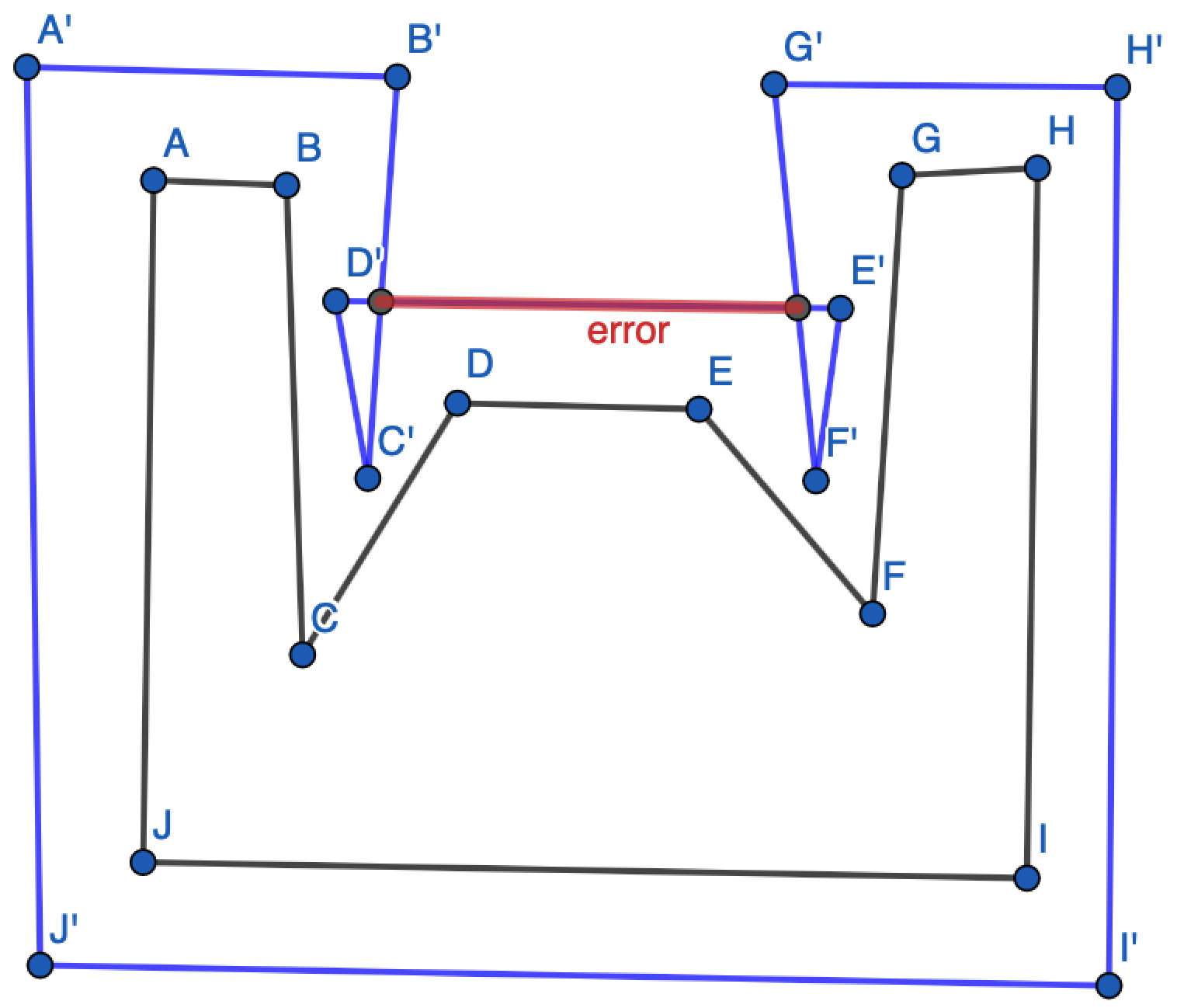}
 \end{minipage}\\
 \caption{An example to show the shortcoming of  direct offsetting method in 2D.
The inner surface(black) and the outer surface (blue) are shown with the outward offsetting operation. The offsetting positions of vertex $D$ and vertex $E$ are   $D^{'}$ and  $E^{'}$ respectively. This case produces self-intersection, and some part of the edge (red) cannot be identified by the searching algorithm.}\label{fig:con0}
\end{figure}

\begin{figure}[htb]
 \begin{minipage}[c]{0.5\textwidth}
    \centering
    \includegraphics[width=3 in]{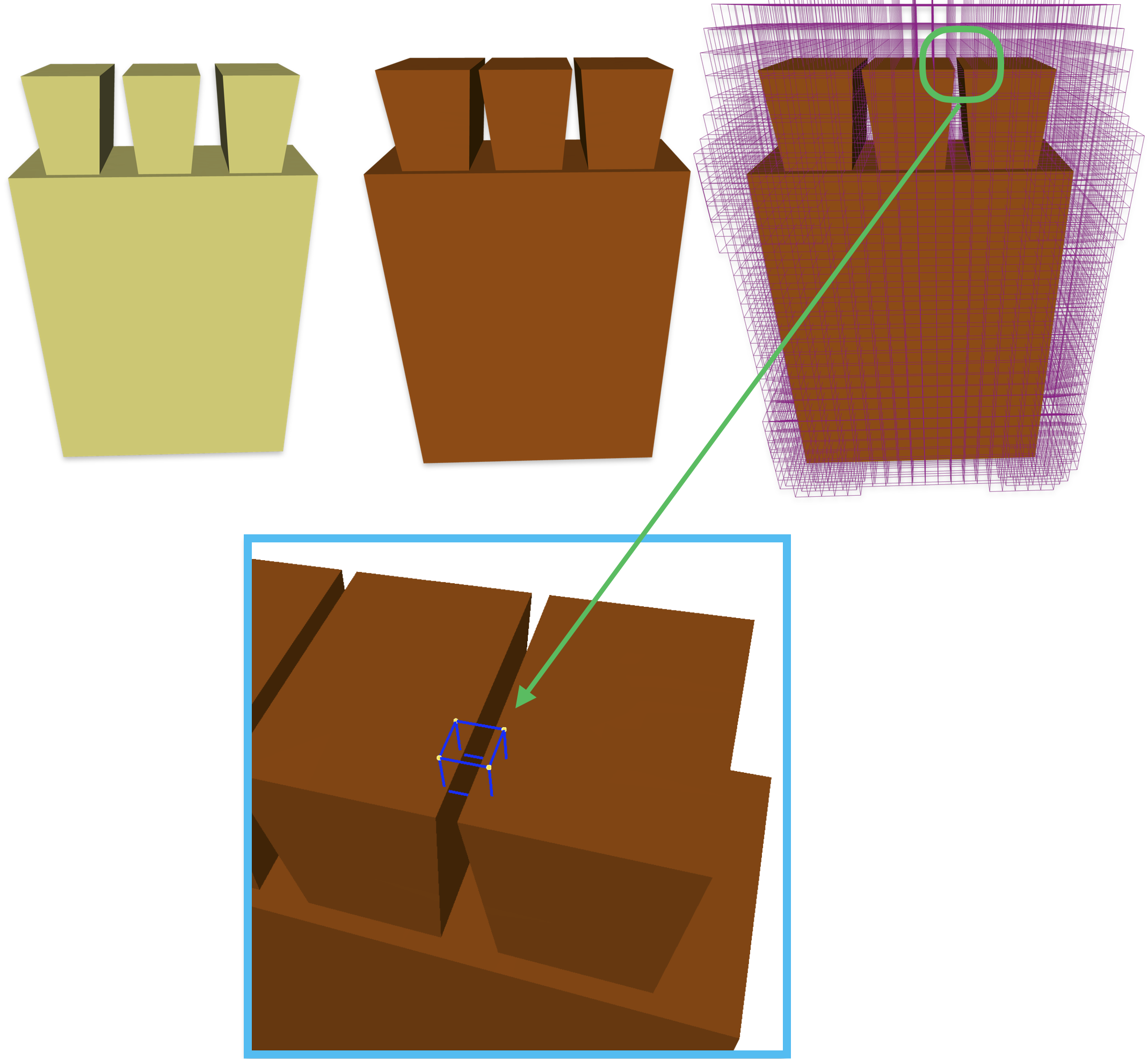}
 \end{minipage}\\
 \caption{In the depicted figure, we present scenarios where the dual contouring mesh generation algorithm encounters challenges. Commencing on the top left, that is the initial mesh. Adjacently in the center-top, that is the resultant mesh derived from specific offsets. On the top right, the sampling grid is illustrated. Expanding upon this, the bottom visualization magnifies a cell from the aforementioned sampling grid, highlighting a particular cell that the dual contouring algorithm fails to accurately discern.}\label{fig:con1}
\end{figure}
\begin{figure}[htb]
 \begin{minipage}[c]{0.5\textwidth}
    \centering
    \includegraphics[width=2.5 in]{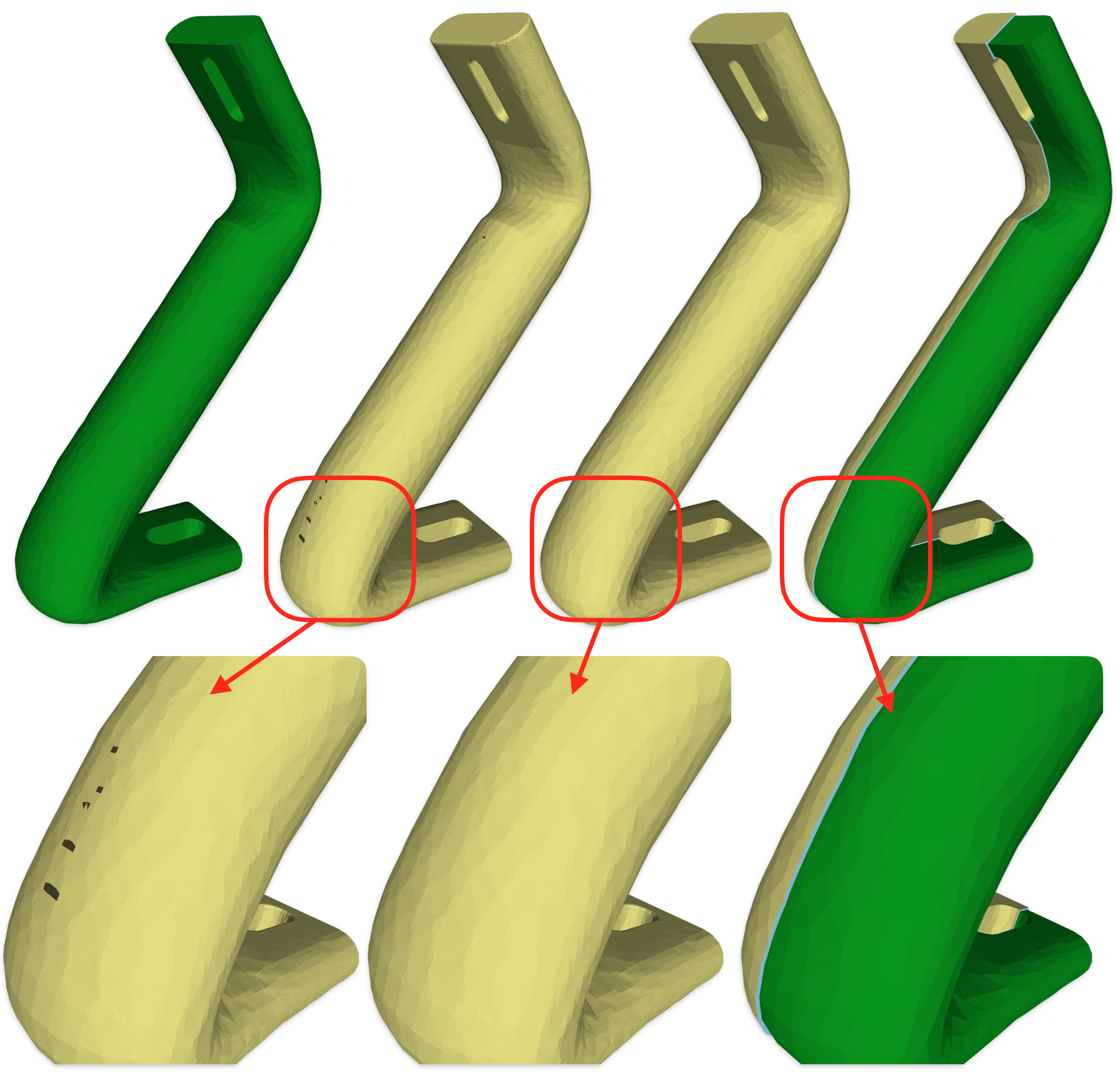}
 \end{minipage}\\
 \caption{An example of the problem caused by sampling when meeting short distance offset. The first mesh from left to right represents the initial mesh, the second represents the result of the sampling method using 256*256*256 outward offset, and the third represents the result of our method's outward offset , the fourth shows the cross-section of our method and put together with the initial mesh.}\label{fig:posun}
\end{figure}

However, currently it is still a challenging problem to create an offset model robustly from the boundary polygon representation. The direct offsetting methods often cause self-intersections as shown in Fig. \ref{fig:con0}, and it is difficult to judge whether to generate after intersection for complex CAD models. The distance-field based methods can accurately define the surface generated after offsetting, but it often causes problems due to the sampling of implicit expressions as shown in Fig. \ref{fig:con1} and Fig. \ref{fig:posun}. Moreover, there is few research work on the feature-preserving and mesh variable offsetting method with high efficiency and robustness.

In this paper, we will focus on the mesh offsetting problem with variable distance. We propose a parallel feature-preserving mesh variable offsetting framework to create an offset model without gaps, holes, and self-intersections. The input is a triangular mesh denoted as $Mesh_0$ and the desired offset distance for each face on $Mesh_0$; The output is the offsetting triangular mesh denoted as $Mesh_1$. Different from the traditional method based on distance and normal vector, a new calculation of offset position is proposed by dynamic programming and quadratic programming. Instead of distance implicit field, a spatial coverage region represented by polyhedral for computing offsets is proposed.

Our offsetting approach has the following several advantages which have not been provided by existing approaches.
\begin{itemize}
    \item \textbf{Variable offsetting}: We provide a practical method to generate offsetting mesh with different distance, which plays an important role for shape optimization results. Arbitrary offset distance can be handled to generate both grown and shrunk models.
    \item \textbf{Feature-preserving}: Distance-field based methods often lack sharp edges or corners indicated in the original models. In contrast, our offsetting results can preserve all the sharp feature of the original model.
    \item \textbf{Efficient}: several acceleration techniques are proposed for the efficient mesh offsetting, such as the parallel computing with grid, AABB tree and rays computing.
    \item \textbf{Light-weight}: Compared with previous method, our framework can generate a offsetting surface with smaller mesh size, which is similar with the original mesh.
\item \textbf{Simple}: Our framework is relatively easy to implement, which can be considered as a combination of direct offsetting method and distance-field based approach. The code of our implementation has been open sourced on Github.
\end{itemize}

%% file: RelatedWork.tex
\section{Related work}
\label{sec:related}

\newcommand{\starzero}{\textcolor{violet}{\times}}
\newcommand{\starone}{\textcolor{red}{\star}}
\newcommand{\startwo}{\textcolor{orange}{\star\star}}
\newcommand{\starthree}{\textcolor{yellow}{\star\star\star}}
\newcommand{\starfour}{\textcolor{lime}{\star\star\star\star}}
\newcommand{\starfive}{\textcolor{green}{\star\star\star\star\star}}

In this section, we discuss various research efforts related to the topic of this paper, spanning methodologies such as direct offsetting, distance-field-based methods, the Minkowski sum, skeleton-based techniques, and ray-based approaches.

\noindent\textbf{Direct Offsetting Methods}

A prominent method proposed by \cite{jung2004self} offsets based on vertex normal direction. Although efficient, it struggles with holes in complex CAD models, leading to offset meshes that can be defective, particularly in scenarios involving self-intersections. The precision issues arising from floating-point operations in self-intersections have been addressed using infinite precision operations \cite{campen2010polygonal}, but results occasionally produce twisted meshes. Offsetting surfaces with polynomials or B-splines is explored in \cite{maekawa1999overview}. However, subsequent intersection operations are intricate. Another approach calculates offset positions based on distance and uniform distribution, followed by point cloud reconstruction \cite{meng2018efficiently}.

\noindent\textbf{Distance-Field-Based Methods}

These methods are popular in modern 3D printing \cite{brunton2021displaced}.
Generally, they require resampling to generate the final mesh, which can compromise geometric features, especially in detailed meshes \cite{wang2013thickening}.
Challenges also arise from the grid density, preservation of sharp features, and computational efficiency.
Some attempts, such as \cite{kobbelt2001feature}, have tackled the ambiguity of the Dual Contouring and Marching Cube, but they tend to excel mostly with CAD models. In the study by \cite{liu2010fast}, there are some improvements in computational efficiency.
The method introduced in \cite{pavic2008high} tries to preserve sharp features, but it involves high grid density and computational expenses.
Only a few works, like \cite{chen2019variable}, delve into variable-thickness offsetting for these methods.

During the generation process, the fixed edge length grid in the dual contouring algorithm is not suitable for all local reconstructions.
An improved method based on octree was proposed in the study by \cite{zint2023feature}.

\noindent\textbf{Minkowski Sum Method}

The Minkowski sum offers a solution for mesh offsetting by calculating the sum of mesh and sphere polygons \cite{rossignac1986offsetting}. An advanced method in \cite{hachenberger2009exact} computes the exact 3D Minkowski sum of non-convex polyhedra by decomposing them into convex parts. Despite its robustness, the method is slow for complex CAD models and struggles with variable thickness offsets and sharp feature preservation. Its implementation in CGAL \cite{cgal:eb-23b} is robust.

\noindent\textbf{Skeleton-Based Methods}

Skeletal meshes are prevalent in geometry processing \cite{tagliasacchi20163d}. The mesh model can be represented by medial axes and spheres \cite{amenta2001power, sun2015medial}. There are techniques to generate skeleton meshes \cite{lam1992thinning, li2015q}. However, they don't always suit all mesh types, and some mesh models have intricate skeletal structures. There have been efforts to simplify medial axes \cite{sun2013medial}. Offsetting can be achieved by adjusting the radius of the balls on the skeleton. Still, these methods often fall short with CAD models with sharp features.

\noindent\textbf{Ray-Based Methods}

This approach \cite{chen2019half, wang2013gpu} is rooted in the dexel buffer structure \cite{van1986real}. A recent algorithm in \cite{chen2019half} offers an efficient parallelized method using rays and voxels to compute mesh offsets, bypassing the need for distance field computation. Nonetheless, due to the grid's voxel-like structure, a high density is essential for accurate shape representation, leading to increased computational costs and dense mesh outputs. The approach also mainly considers a consistent offset, and the results are usually voxel representations.

\begin{table*}[htb]
    \centering
    \begin{tabular}{p{4cm}p{1.55cm}p{1.55cm}p{1.45cm}p{1.55cm}p{1.55cm}p{1.55cm}p{1.55cm}}
        \toprule
        \textbf{} & \textbf{Support Variable} & \textbf{Efficiency with short distance} & \textbf{Efficiency with long distance} & \textbf{Sharp feature preserving} & \textbf{Ambiguity} & \textbf{Robustness}  & \textbf{mesh/voxels}\\
        \midrule
        Direct offset method\cite{jung2004self} & $\starfive$ & $\starfive$ & $\starfour$ & $\starthree$ & $\starfive$ & $\startwo$ & mesh\\
        Distance field\cite{wang2013thickening} & $\starone$ & $\starthree$ & $\starfive$ & $\startwo$ & $\starthree$ & $\starthree$ & mesh\\
        Distance field\cite{chen2019variable} & $\starfive$ & $\startwo$ & $\starfour$ & $\startwo$ & $\starthree$ & $\starthree$ & mesh\\
        Minkowski sum\cite{hachenberger2009exact} & $\starone$ & $\starfour$ & $\starfour$ & $\starthree$ & $\starfive$ & $\starfive$ & mesh\\
        Ray based method\cite{chen2019half} & $\starone$ & $\starfour$ & $\starfive$ & $\starthree$ & $\starfive$ & $\starfive$ & voxels\\
        Ours & $\starfive$ & $\starthree$ & $\startwo$ & $\starfive$ & $\starfive$ & $\starfive$ & mesh\\
        \bottomrule
    \end{tabular}

    \caption{Compared with the existing methods, the more the number of " $\star$" in each cell, the better the performance.}
    \label{tab:diff}
\end{table*}

%% file: Method.tex
\section{Method}
\label{sec:Method}

\begin{figure*}[htb]
\begin{minipage}[c]{1.0\textwidth}
    \centering
    \includegraphics[width=6 in]{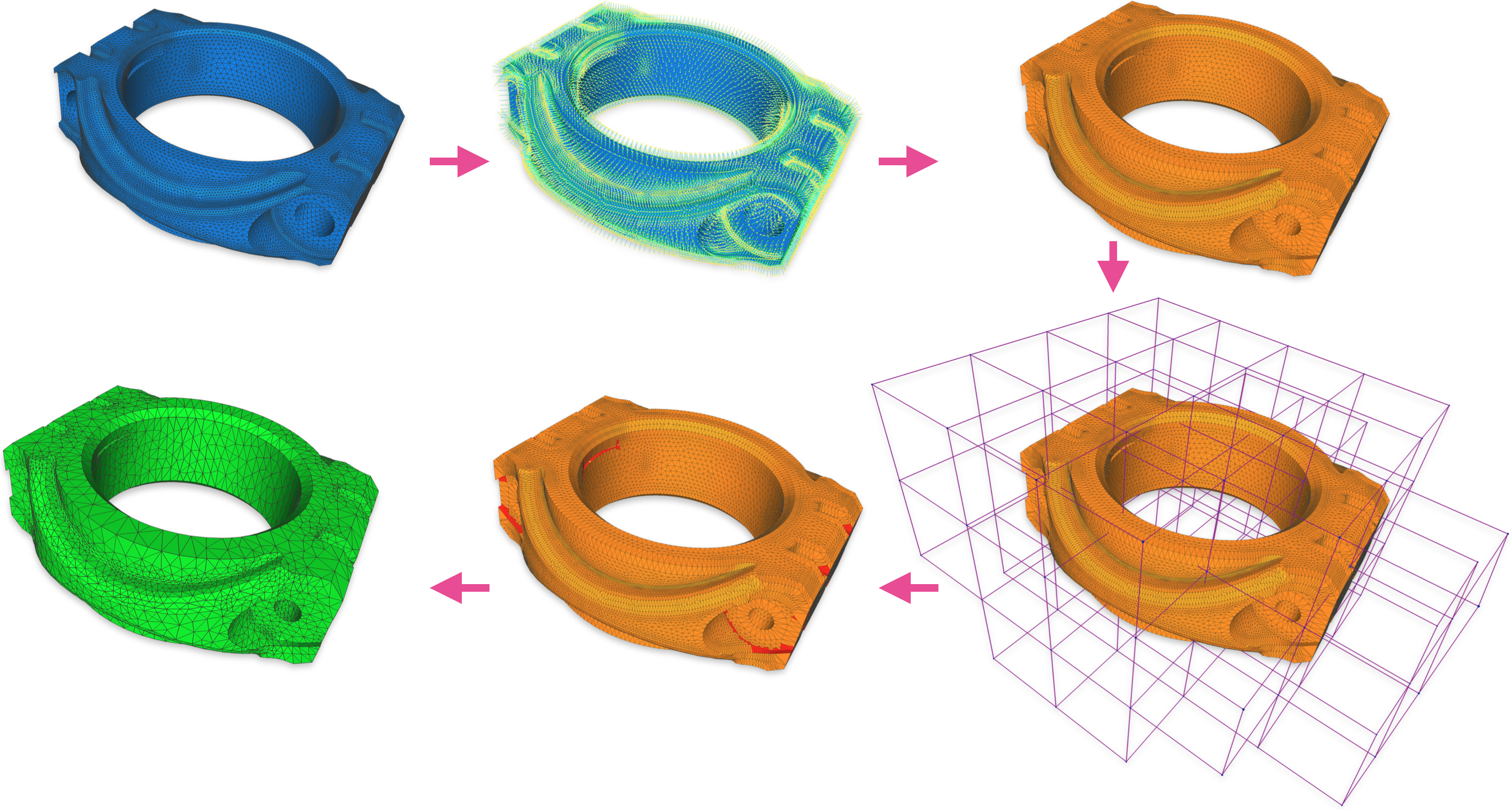}
 \end{minipage}%
 \caption{
 This figure shows the flow of our method, which consists of 6 parts arranged by 5 pink unidirectional arrows.
     The first part is the input mesh.
     The second part is the result of Step $S_1$, showing the path formed by each vertex offset.
     The third part represents the calculated cumulative spatial coverage.
     The fourth part represents the established grid.
     The fifth part represents the calculation of intersection.
     The sixth part is post-processing.
      }\label{fig:mainf1}

\end{figure*}

\begin{figure}[htb]
 \begin{minipage}[c]{0.5\textwidth}
    \centering
    \includegraphics[width=3.3 in]{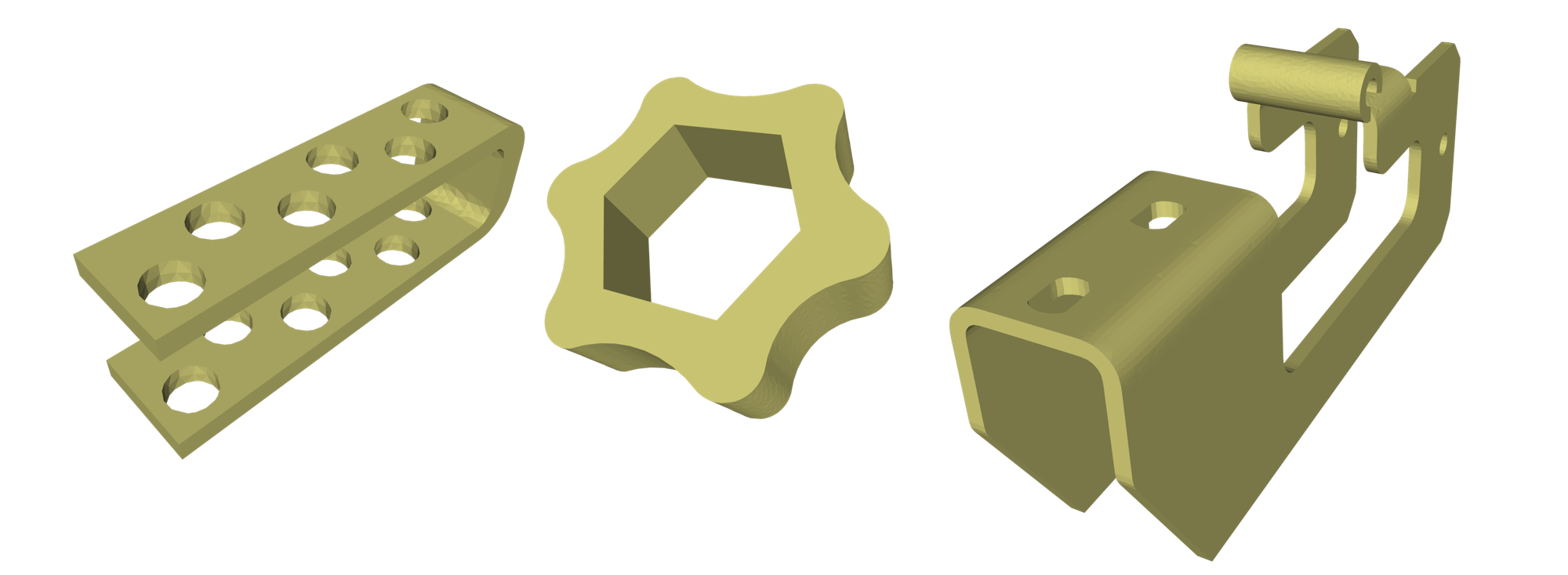}
 \\   (a)
 \end{minipage}\\
 \begin{minipage}[c]{0.5\textwidth}
   \centering
   \includegraphics[width=3.3 in]{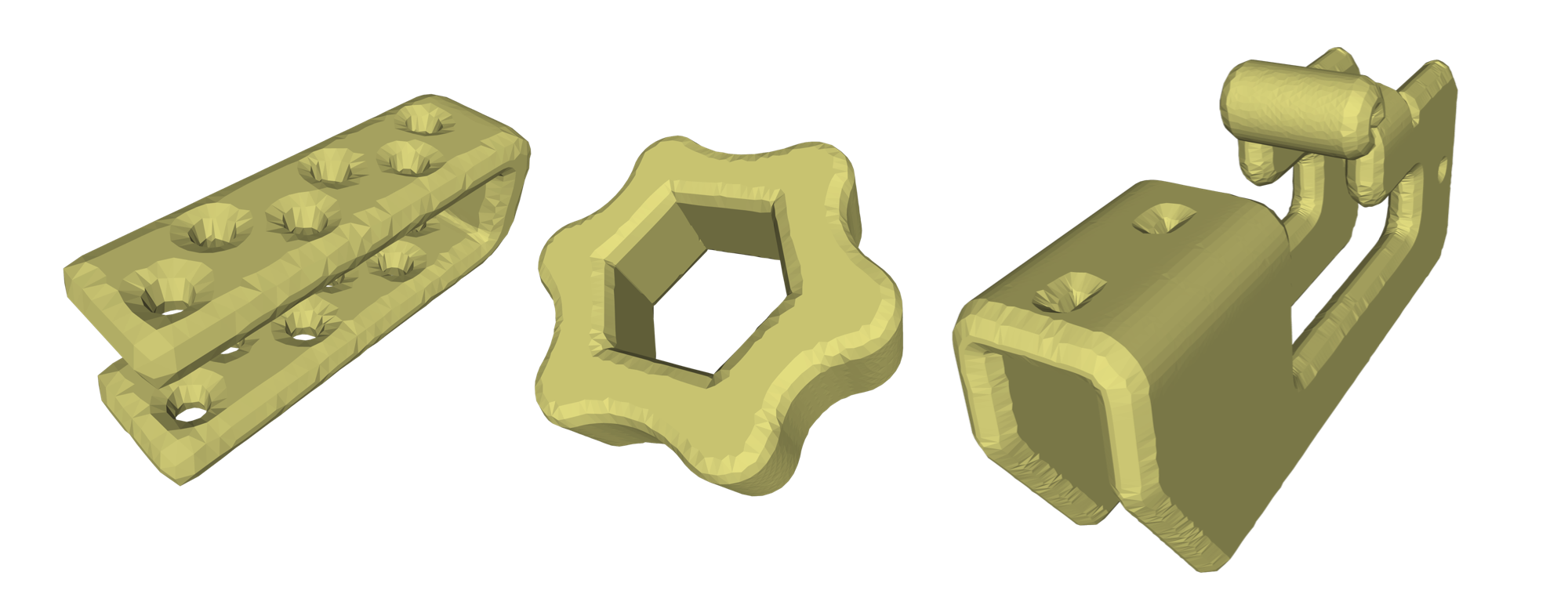}
   \\ (b)
  \end{minipage}\\
  \begin{minipage}[c]{0.5\textwidth}
   \centering
   \includegraphics[width=3.3 in]{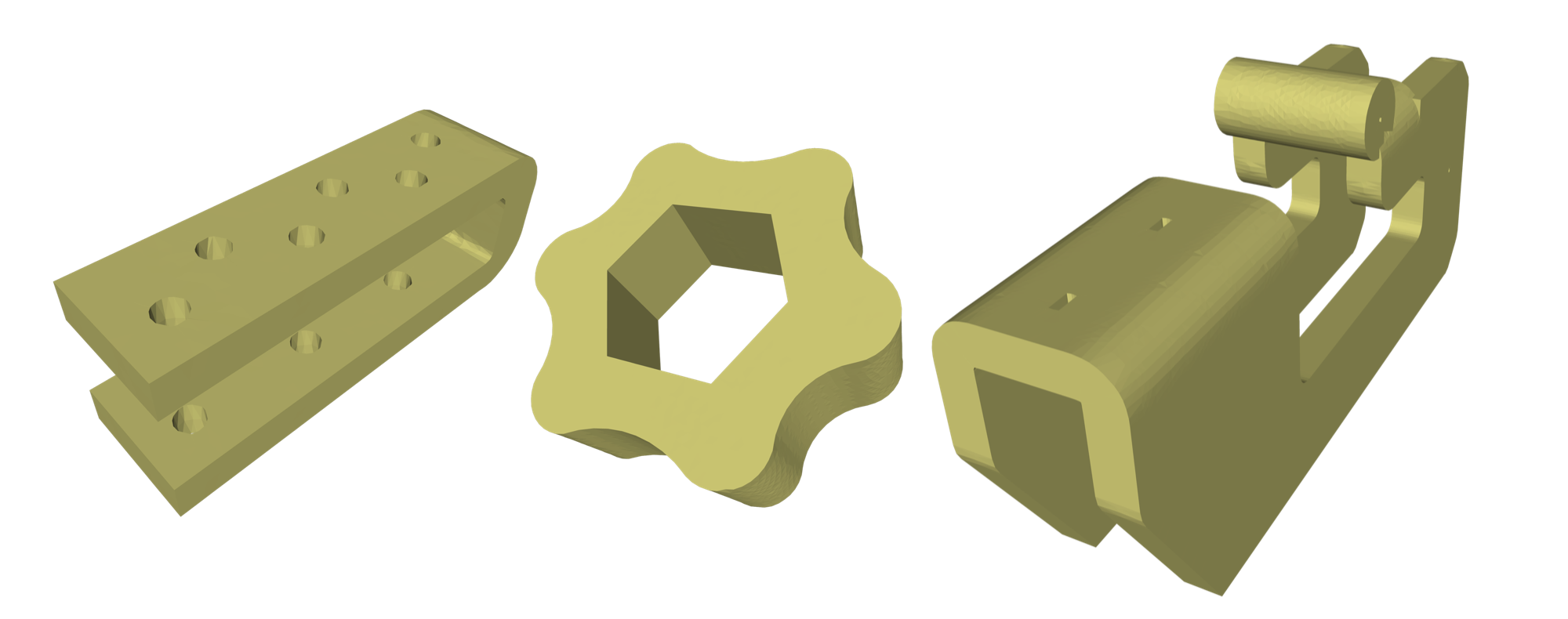}
   \\ (c)
  \end{minipage}\\
 \caption{(a) is origin surface mesh.
     (b) is the result surface mesh through each vertex offsetting outward with normal direction.
    (c) represents our desired mesh offset outcome while preserving sharp features. Therefore, just offsetting based on the normal vectors of each facet does not align with our objective.
    }\label{fig:s1f0}
\end{figure}

\begin{figure}[htb]
 \begin{minipage}[c]{0.5\textwidth}
    \centering
    \includegraphics[width=2.5 in]{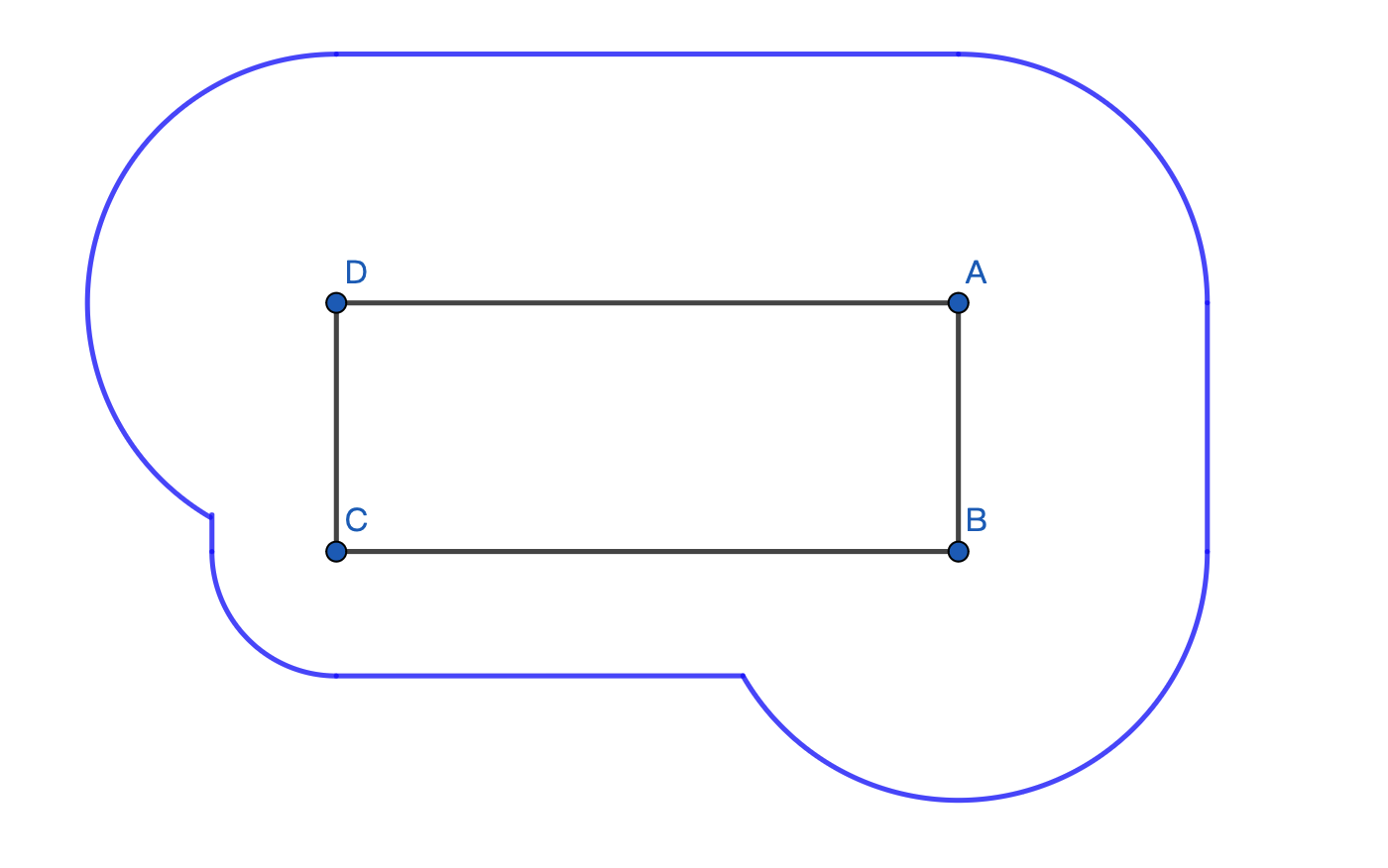}
 \\   (a)
 \end{minipage}\\
 \begin{minipage}[c]{0.5\textwidth}
   \centering
   \includegraphics[width=2.5 in]{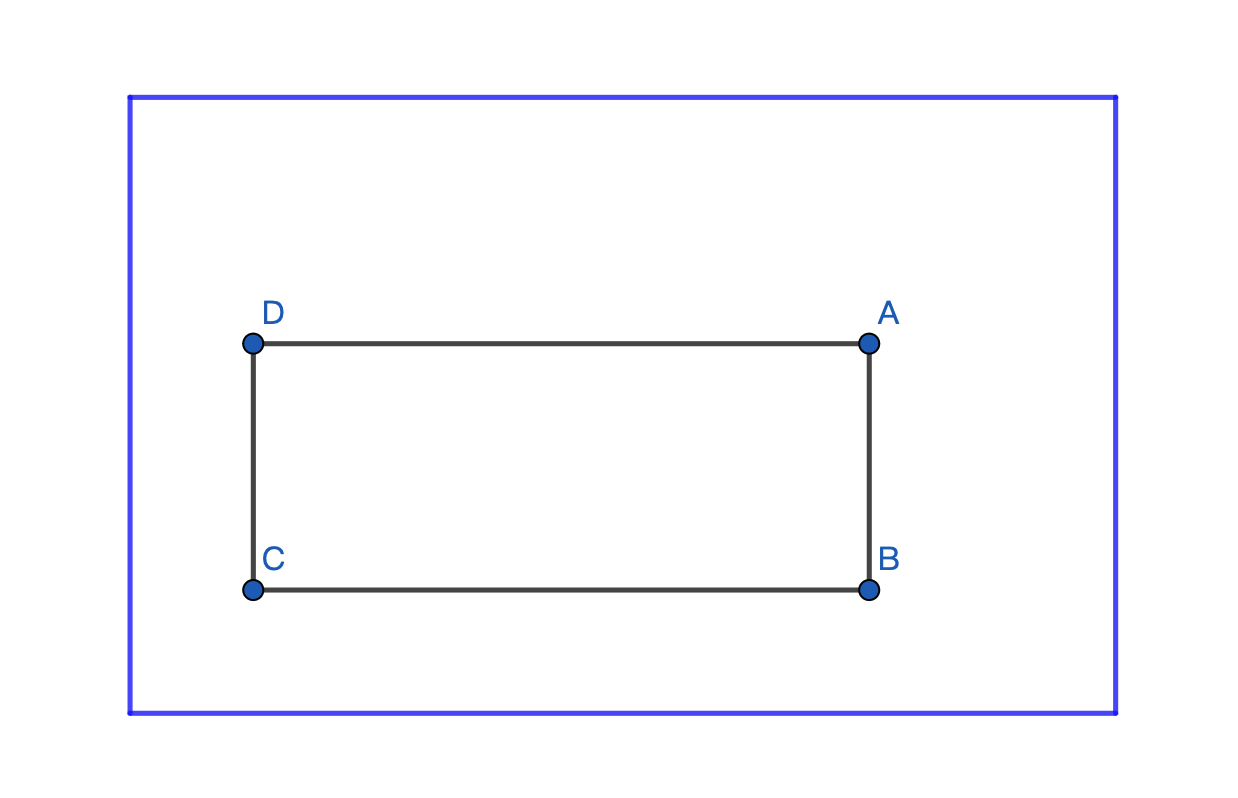}
   \\ (b)
  \end{minipage}\\
 \caption{
  In the figure, the rectangle formed by points $A$, $B$, $C$, and $D$ represents the initial mesh. The offsetting distance of lines $AB$ and $AD$ is $2$, the offsetting distance of lines $BC$ and $CD$ is $1$.
 (a) represents the offset result based on the distance field. (b) represents the desired offset result we aim for, preserving sharp features while performing variable offsetting. Both of which are schematic diagrams of rectangle offsetting in the 2-dimensional case. However, it can be observed that the distance field based approach does not fulfill our goal of preserving features.
 }\label{fig:s1f1}
\end{figure}

We initiated our exploration from two foundational methodologies.
The first involves the direct offsetting method, where every mesh vertex is offset according to its normal vector, either inwards or outwards.
This method yields results characterized by arc-shaped features, as depicted in Fig. \ref{fig:s1f0}(b).
Next, we considered the method rooted in distance fields.
As illustrated in Fig. \ref{fig:s1f1}(a), relying on distance fields for offset calculation would still manifest arc-shaped or spherical features.
Despite its challenges in preserving sharp features, it uniquely delineates a region within a three-dimensional space.
This defined region corresponds to the spatial coverage resulting from the offset operation.
Robustness in various complex scenarios is achieved by resampling the boundaries of this region.

Both aforementioned methods fall short in retaining sharp features.
However, upon closer inspection, the direct offsetting method presented opportunities for refinement.
Adjustments in the offset's magnitude and direction can potentially aid in the preservation of sharp features.
To capitalize on the robustness inherent to the distance field method, our proposed approach represents the offset-induced spatial coverage through the union of multiple polyhedra. This integrated approach has the dual advantages of sharp feature preservation and robustness in intricate models.

For every face, \(face_i\), of a triangular mesh, \(Mesh_0\), the offset results in a polyhedron.
This polyhedron represents the spatial coverage of \(face_i\), and is denoted as \(F_i\).

Given that \(Mesh_0\) consists of \(n\) triangular faces, the cumulative spatial coverage post-offset is represented by \(\Sigma_{i=1}^{n}{F_i}\). This coverage manifests as a shell-like structure with an outer surface and an inner surface corresponding to a hollow body. In the context of an outward offset, \(Mesh_0\) is defined by the inner surface, while for an inward offset, \(Mesh_0\) corresponds to the outer surface. \(Mesh_0\) is determined based on the offset direction, our subsequent endeavors concentrate on extracting the opposing surface to form \(Mesh_1\).


Essentially, this challenge can be rephrased as the task of merging all \(F_i\) into a hollow polyhedron.
Initial efforts involved leveraging CGAL's polygonal Boolean operations, But we encountered program anomalies after performing a significant number of $F_i$ unions.
Subsequent endeavors utilizing a bespoke Boolean operation proved insufficient in terms of robustness and efficiency.

Consequently, we adopted an approach that momentarily disregards topological relationships, focusing on intersection, subdivision, followed by ray probing techniques.
This method produced a Triangle Soup constituting the offset surface.
Subsequently, we employed TetWild to transform this into tetrahedra, extracting the external surface to realize the offset surface.
Although this methodology may cede some efficiency, it boasts commendable robustness.
Even if minor errors (less than 0.01\%) arise in the Triangle Soup due to discrepancies in previous steps, the approach remains effective in delivering robust results.

The subsequent subsections will describe details of our methodology.
Our approach is articulated through five steps, designated as \(S_1 \sim S_5\).
Detailed descriptions of these steps are provided from subsection \ref{subsec:s1} to subsection \ref{subsec:s5}, respectively.

%
%
%
%
%
%

\input{4.1.tex}

\input{4.2.tex}
\input{4.3.tex}
\input{4.4.tex}
\input{4.5.tex}

%% file: 4.1.tex
\subsection{Dynamic Programming and Quadratic Programming}
\label{subsec:s1}

This step is a pivotal one in our approach.
It incorporates both dynamic programming and quadratic programming to determine the optimal offset position for each vertex in \(Mesh_0\).
Initially, our offsetting procedure originates from individual vertices, contemplating a singular offset position for each vertex.

The preservation of features can be independently addressed by considering local meshes.
A vertex \(V\) from \(Mesh_0\), in conjunction with several faces from its 1-ring neighborhood, constitutes a local mesh.
Given \(k\) faces within this local mesh, we denote each face as \(O_i\) (\(0 \le i \le k-1\)).
The desired offset distance for \(O_i\) is \(L_i\), and the plane created by \(O_i\) is represented as \(A_i\).

To maintain sharp features while performing a variable offset on the mesh, it becomes evident that distance cannot serve as a metric of feature-preserving due to the varied offset distances across facets. Instead, employing angles as a metric is a viable solution. Extensive testing indicated that for the offset vertex \(V'\) derived from \(V\), in order to preserve the angular attributes of the local mesh, the distance from \(V'\) to the plane \(A_i\) should be equivalent to \(L_i\).

To circumvent numerical precision issues that might arise during subsequent solver computations, we introduce two tolerance coefficients, \(tole_l\) and \(tole_r\).
Hence, the offset distance condition can be relaxed, ensuring the offset distance falls within the range $tole_lL_i \sim tole_rL_i$.
The two tolerance coefficients can be adjusted by the user. By default, our software sets \(tole_l = 1.0\) and \(tole_r = 1.3\).

In instances where there are infinite points that can serve as \(V'\) satisfying the aforementioned distance constraints, for example, when all facets in the 1-ring neighborhood of \(V\) lie on the same plane. Our experiments reveal that selecting the point closest to \(V\) as \(V'\) within this infinite solution space achieves the best feature fidelity.
So, the optimization objective strives to minimize the distance between \(V'\) and \(V\).
Considering the quadratic nature of point-to-point distances juxtaposed with a set of linear constraints defined by the point-to-plane distances, our problem conforms to a quadratic programming framework.

The formulation of this quadratic programming equation proceeds as follows:

The distance from $V^{'}$ to $V$ is $D$.
The coordinate of $V^{'}$ is $(x, y, z)^{T}$ abbreviated as $\vec{x}$.
The coordinate of $V$ is $(x_p,y_p,z_p)$.

\begin{equation}
D=(x-x_p)^2+(y-y_p)^2+(z-z_p)^2
\end{equation}

The formula can be deduced as

\begin{subequations}
\begin{align}
D=\vec{x}^{T}\begin{bmatrix} 1 & 0 & 0 \\0 & 1 & 0 \\0 & 0 & 1 \\\end{bmatrix}\vec{x} - 2e_1\vec{x} + e_2 \\
e_1 = x_p+y_p+z_p \\
e_2 = x_p^2 +y_p^2+z_p^2
\end{align}
\end{subequations}

The random point on plane $A_i$ is $(x_i,y_i,z_i)^T$.
The normal vector of plane $A_i$ is $(nx_i,ny_i,nz_i)^T$ abbreviated as $\vec{n_i}$.
The number of faces in 1-ring neighborhood of $V$ is $k$.
Constraints are established according to the distance $L_i$ from $V^{'}$ to plane $A_i$:

\begin{equation}
tole_l L_i \le\frac{(x-x_i,y-y_i,z-z_i)\vec{n_i}}{|\vec{n_i}|} \le tole_r L_i,0 \le i \le k-1
\end{equation}

Constraints can be deduced as
\begin{equation}
tole_lL_i|\vec{n_i}|+(x_i,y_i,z_i)\vec{n_i} \le \vec{n_i}^T\vec{x} \le tole_rL_i|\vec{n_i}|+(x_i,y_i,z_i)\vec{n_i}
\end{equation}

set $\overrightarrow{min_i} = tole_l L_i|\vec{n_i}|+(x_i,y_i,z_i)\vec{n_i}$, and $\overrightarrow{max_i} = tole_r L_i|\vec{n_i}|+(x_i,y_i,z_i)\vec{n_i}$.

In optimizing goal $D$, $e_2$ has no effect, so $e_2$ is discarded.
Then the final planning goal is:
\begin{subequations}
\begin{align}
D=\vec{x}^{T}\begin{bmatrix} 1 & 0 & 0 \\0 & 1 & 0 \\0 & 0 & 1 \\\end{bmatrix}\vec{x} - 2e_1\vec{x} \\
\overrightarrow{min_i}  \le \vec{n_i}^T\vec{x} \le \overrightarrow{max_i},0 \le i \le k-1
\end{align}
\end{subequations}

This can be resolved using a quadratic programming solver, such as OSQP \cite{osqp}.
However, certain scenarios render the above equations unsolvable, for instance, mesh vertices in slender dihedral angles or vertices shared by regions with opposing normal vectors.
To address it, we propose a mechanism where a individual mesh vertex can yield multiple offset vertices.
This technique necessitates the use of state compression dynamic programming, which will be elaborated upon in subsequent portion of this subsection.

%

\begin{figure}[htb]
 \begin{minipage}[c]{0.5\textwidth}
    \centering
    \includegraphics[width=2.4 in]{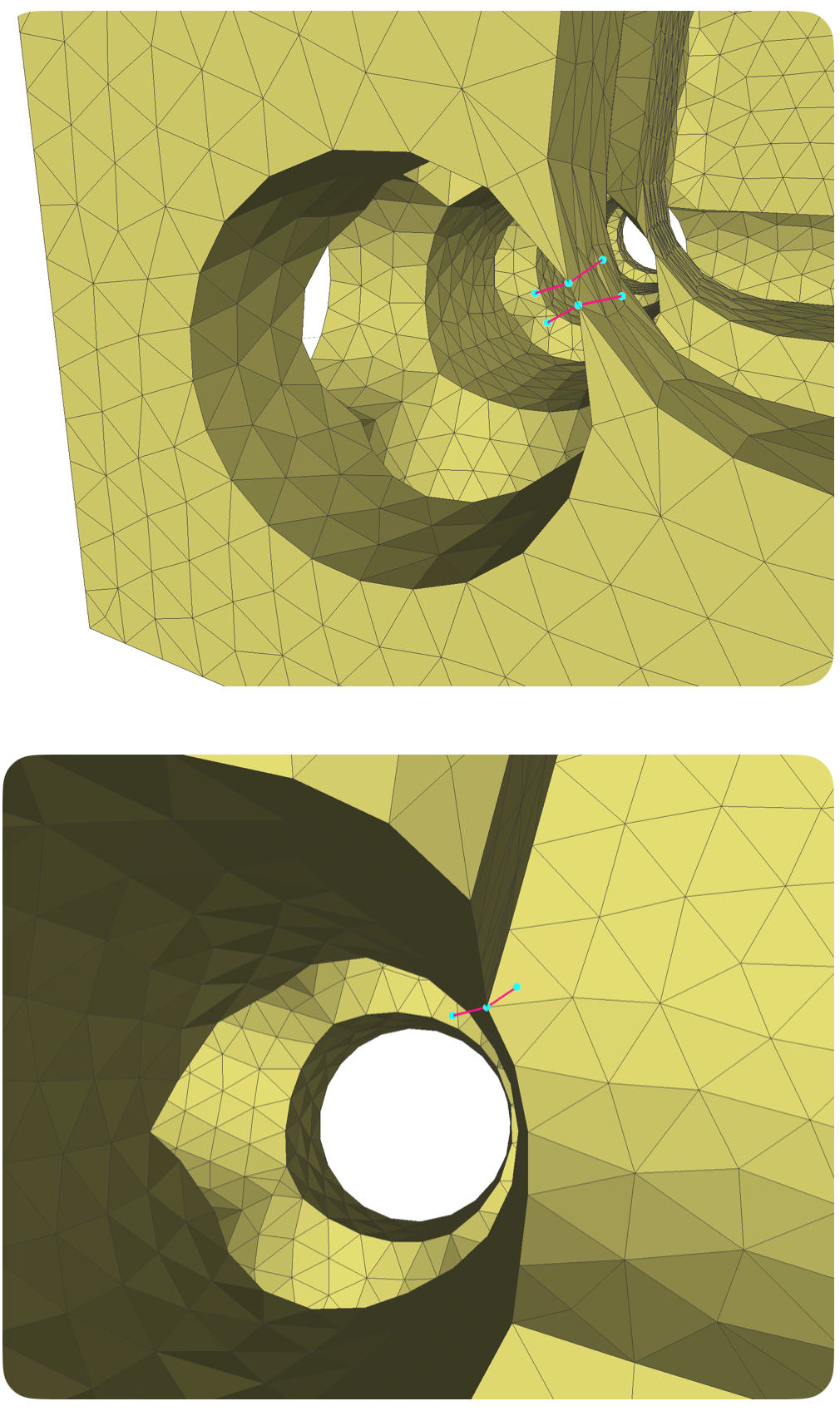}
 \end{minipage}\\
 \caption{In our method, some vertices when subjected to a mere positional displacement from one locale to another, each of these vertices engenders infeasibilities within quadratic programming.
Our solution illustrated in this figure is manifested by displacing each of these vertices from its incumbent position to two distinct positions, yielding a pair of offset vertices.
To facilitate the derivation of two or more such offset points, this study harnesses a state-based dynamic programming approach. }\label{fig:s2023s41}
\end{figure}

There exist scenarios where quadratic programming lacks feasible solutions. In such cases, our approach, as depicted in the Fig. \ref{fig:s2023s41}, involves displacing certain vertices from their incumbent positions to two distinct positions, resulting in a pair of offset points. This technique enables the derivation of two or more such offset points and is facilitated by a state-based dynamic programming approach.

We have refined the aforementioned approach by first categorizing each \(O_i\) into groups, totaling \(m\) distinct groups.
Each group comprises a certain number of triangular facets, and each \(O_i\) exclusively appears in one of these groups.
The \(r^{th}\) group contains \(n\) triangular facets, denoted as \(k_{r1}\), \(k_{r2}\), and so on.
Using the planes generated by \(k_{r1},k_{r2},\dots k_{rn}\) and applying the aforementioned quadratic programming method, an offset vertex of the \(r^{th}\) group , \(V_r^{'}\), is derived.

Given that the local mesh is categorized into \(m\) groups, our optimization objective is set as \(D_{NEW} = \sum_{r=1}^{m}||V-V_r^{'}||^2\). This choice is based on extensive experimentation, as we have observed that minimizing $D_{NEW}$ results in better preservation of the mesh's features.

A pivotal question arises: How can these facets be optimally grouped into \(m\) categories to minimize the target \(D_{NEW}\)? Further, how is this $m$ determined? However, it is worth noting that implementing a method based solely on grouping facets by their normal vectors may lack robustness. To tackle this, we adopt a state-based dynamic programming approach.

We employ a state-based dynamic programming approach for implementation.
In this context, there are \(k\) facets, with each facet having two potential states: "selected" and "unselected".
As a result, there are a total of \(2^k\) possible combination states.
We utilize a \(dp\) array with a capacity of \(2^k\) to store the \(D_{NEW}\) values corresponding to each combination of facet selections.
The binary representation \(0b010101\) is used to denote the number whose binary value is \(010101\).
For \(n=6\), the binary \(0b010101\) signifies the selection of the first, third, and fifth facets, whereas \(0b100100\) indicates the selection of the third and sixth facets.

Consider a current state \(state_0\).
If the combination's facets, upon inspection, yield an unsolvable quadratic programming problem, they can be decomposed into two sets: \(state_1\) and \(state_2\), satisfying \(state_1 \cup state_2 = state_0\) and \(state_1 \cap state_2 = \emptyset\).
The value \(dp[0b111111]\) represents the best solution from all divisions.
Our primary aim isn't minimizing the number of splits, but ensuring that the cumulative quadratic programming objective \(D\) across all sub-combinations, \(D_{NEW}\), is minimized. When \(D_{NEW}\) reaches its minimum, the number of split groups corresponds to \(m\).
A recursive approach is used for solution, illustrated in the Algorithm \ref{alg:1}.

\begin{algorithm}
    \SetAlgoLined 
    \caption{State-based Dynamic Programming}
    \label{alg:1}
    \KwIn{Vertex V, 1-ring neighbor facet's list}
    \KwOut{The minimum value of D\_NEW}

    $k \gets$ The size of 1-ring neighbor facet's list\;
    Initialize vectors $dp$ of size $2^{k}$\;
    Fill $dp$ with $-1$\;
    Initialize empty vectors $sub\_state$\;
    \SetKwProg{Fn}{Function}{:}{end}

    \Fn{get\_sub\_state(state,pos,change)}{
        \If{$pos = -1\ and\ change = True$}{
            Put $state$ at the end of $sub\_state$\;
            \Return{\ }
        }
        \For{$i \gets pos\ to\ 0$}{
            \If{$state \& (2^i) \neq 0$}{
                \textit{get\_sub\_state($state,i-1,change$)}\;
                \textit{get\_sub\_state($state \oplus (2^i),i-1,True$)}\;
                \textbf{break}
            }
            \Else{
                 \If{$i = 0$}{
                    \textit{get\_sub\_state($state,i-1,change$)}\;
                    \textbf{break}
                }
            }
        }
    }

    \Fn{DFS(state)}{
        \If{$dp[state] \neq -1$}{
             \Return{$dp[state]$}
        }
        $success, V_p \gets \textit{do\_quadratic\_programming}(state)$\;
        \If{$success = True$}{
            $dp[state] \gets$ $sqrt(||V_p-V||^2)$\;
            \Return {$dp[state]$}
        }
        clear the vector  $sub\_state$\;
        \textit{get\_sub\_state($state,k,False$)}\;
        $sub\_state\_tmp \gets sub\_state$\;
        $minx \gets \infty$\;
        \For{$next\_state \in sub\_state\_tmp$}{
            $sub\_ans \gets$ \textit{DFS($next\_state$)} + \textit{DFS($state - next\_state$)}\;
                \If{$sub\_ans < minx$}{
                    $minx \gets sub\_ans$\;
                }
        }
        $dp[state] \gets minx$\;
        \Return {$minx$}
    }
    \textit{DFS($(2^k)-1$)}

\end{algorithm}

The computational complexity of this dynamic programming strategy is \(O(3^k + 2 - 2 \times 2^k )\). When \(k=1,2,3,4,5,6,7,...\), the resulting sequence is \(1, 3, 13, 51, 181, 603, 1933,...\), which is identified as sequence A101052 on the OEIS website [https://oeis.org/]. Mathematical derivations pertaining to such issues can be found in \cite{giraudo2015combinatorial}.

For \(k \leq 16\), the computational iterations are fewer than \(42915651\).
Consequently, when input mesh regions exhibit a vertex which 1-ring neighbor facet counts of 17 or greater, it's beneficial to implement a triangle mesh remeshing algorithm prior to our method.
\begin{figure}[htb]
 \begin{minipage}[c]{0.5\textwidth}
    \centering
    \includegraphics[width=2.2 in]{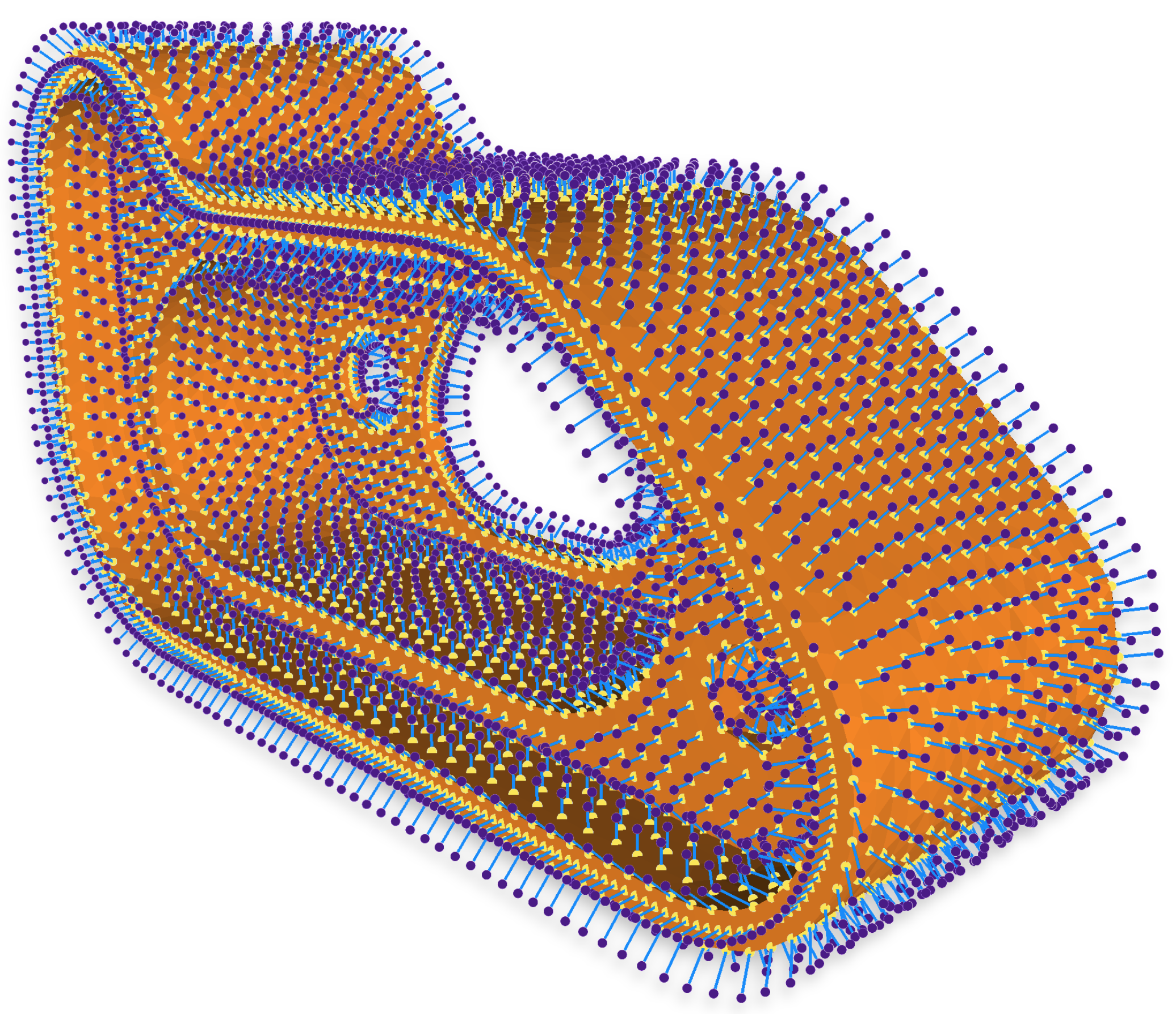}
 \end{minipage}\\

 \caption{In the figure, a mesh is depicted, with each yellow point representing an individual vertex of this mesh.
 The purple points denote the newly generated vertices resulting from the offset applied to these original vertices.}\label{fig:31}
\end{figure}

After each vertex in \(Mesh_0\) has undergone the aforementioned dynamic programming process, the result will resemble what is depicted in Fig. \ref{fig:31}. Additional, in the above approach, further optimization can be achieved by recursively decomposing the current state into smaller substates, even when quadratic programming at the current state has feasible solutions. A comparison is then made between the offset distance of the feasible solution at the current state and the results obtained from the recursive decomposition to determine which one results in a shorter total offset distance.

%% file: 4.2.tex
\subsection{Reassign by Octree}
\label{subsec:s2}

In subsection \ref{subsec:s1}, the generation of each vertex primarily takes into account the local mesh characteristics in isolation, without considering inter-vertex relationships.
Consequently, in certain specialized scenarios, a proliferation of closely situated vertices can manifest.

To ameliorate mesh quality and reduce subsequent computational complexity, we opt to merge closely located vertices into a singular vertex within our approach.
Specifically, when the Euclidean distance between two vertices derived from offset calculations is less than $\varepsilon$, these two vertices are consolidated into a unified entity.

%
%
%

Nonetheless, this process may engender a cascade effect of mergers, as exemplified by the merger of ${V_1}^{'}$ with ${V_2}^{'}$, ${V_2}^{'}$ with ${V_3}^{'}$, and ${V_3}^{'}$ with ${V_4}^{'}$, resulting in the eventual merger of ${V_1}^{'}$ and ${V_4}^{'}$ due to transitive relationships.

In instances where the Euclidean distance between ${V_1}^{'}$ and ${V_4}^{'}$ is relatively large, their merger can induce localized distortions within the result mesh.
To address this challenge, we have devised an approach that entails an initial merging phase followed by a subsequent splitting phase.

\begin{figure}[htb]
 \begin{minipage}[c]{0.5\textwidth}
    \centering
    \includegraphics[width=2.8 in]{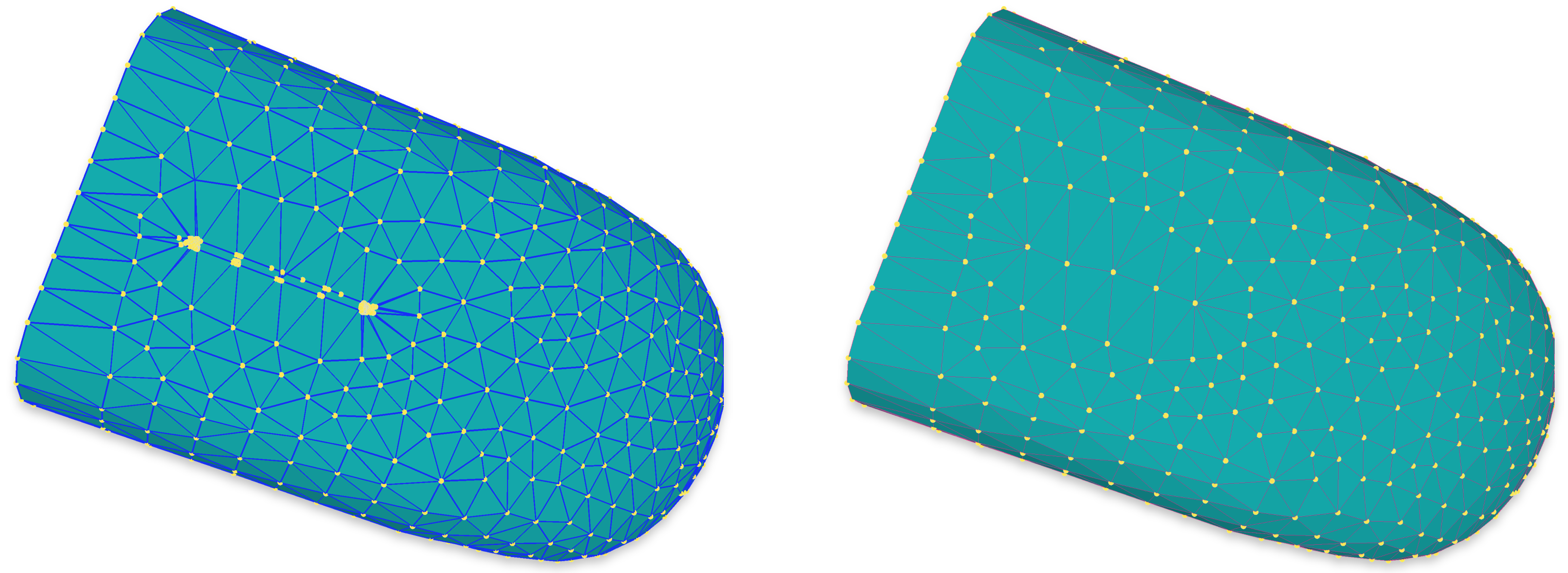}
 \end{minipage}\\

 \caption{ The figure is bifurcated into two distinct sections, each illustrating the cumulative spatial coverage constructed during Step \(S_3\).
 The left section represents the scenario where Step \(S_3\) is executed directly without Step \(S_2\), while the right section depicts the sequence in which Step \(S_2\) is executed prior to Step \(S_3\).}\label{fig:s1nf5}
\end{figure}

This entails employing an octree structure to reassign each merged vertex collection into multiple discrete points (see Algorithm \ref{alg:2} for further details). The impact of this step on the subsequent step (Step \(S_3\)) is illustrated in Fig. \ref{fig:s1nf5}.



\begin{algorithm}
    \SetAlgoLined 
    \caption{Reassign Programming By Octree}
    \label{alg:2}
    \KwIn{A set of vertices generated by merging}
    \KwOut{The result of reassign}
    $v\_list[root] \gets$\ The\ set\ of\ vertices\ generated\ by\ merging\;
    \SetKwProg{Fn}{Function}{:}{end}

    \Fn{reassign(node)}{
        $ max\_d \gets 0$\;
        $l \gets size(v\_list(node))$\;
        \For{$i \gets 0\ to\ l-1$}{
            \For{$j \gets 0\ to\ l-1$}{
                $d \gets distance(v\_list[node][i],v\_list[node][j])$\;
                $max\_d = \max(max\_d,d)$\;
            }
        }
        \If{$max\_d < limit$}{
            $p \gets$\ The\ average\ coordinate\ of $v\_list[node]$\;
            \For{$i \gets 0\ to\ l-1$}{
                $v\_list[node][i] \gets p$\;
            }
            \Return{\ }
        }
         \For{$i \gets 0\ to\ l-1$}{
            $child_x \gets $The\ child\ node\ where $v\_list[node][i]$\ located\;
            append\ $v\_list[node][i]$\ to\ $v\_list[child_x]$\;
         }
        \For{$i \gets 0\ to\ 7$}{
            \textit{reassign($child_i$)}\;
        }
    }

    \textit{reassign(root)}\;

\end{algorithm}

%% file: 4.3.tex
\subsection{Construct the Spatial Coverage of Each Face in $Mesh_0$}
\label{subsec:s3}

Once all three vertices of a triangular facet have undergone the aforementioned computation, we can then delineate its corresponding spatial coverage.
This process entails conducting a Delaunay tetrahedralization using the triangle's three vertices combined with their corresponding offset vertices.

\begin{figure}[htb]
 \begin{minipage}[c]{0.5\textwidth}
    \centering
    \includegraphics[width=3 in]{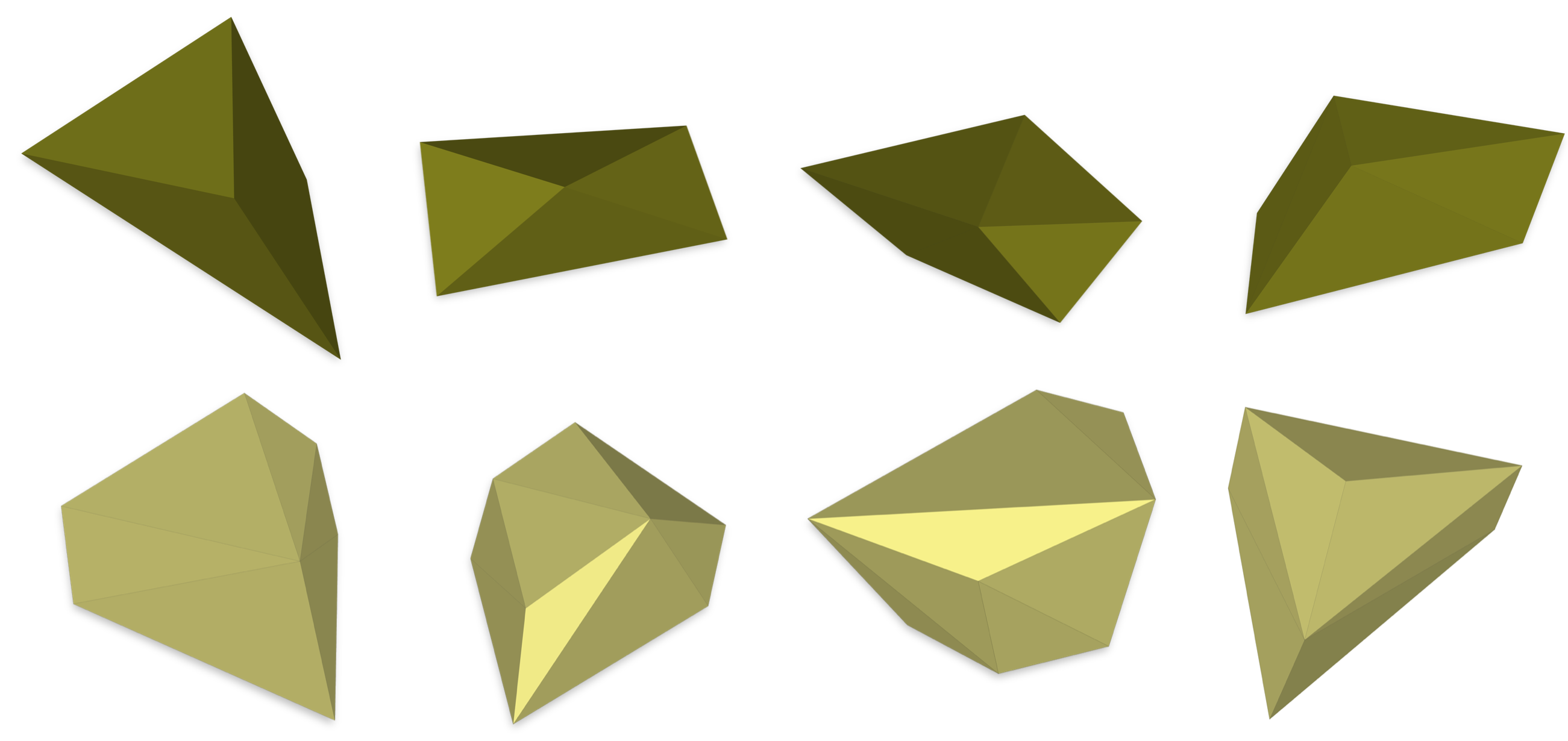}
 \end{minipage}\\
 \caption{Within the displayed figure, the polyhedra in the first row represent the spatial coverage derived when all vertices are offset to an alternative position.
Conversely, the second row illustrates the spatial coverage formulated when certain vertices offset to two or more distinct position.
 }\label{fig:s2f2}
\end{figure}

Next, the exterior surface of the Delaunay tetrahedralization result is computed. The volume enclosed by this exterior surface denotes the spatial coverage of the corresponding facet. For a given facet \(face_i\), its computed spatial coverage is denoted by \(F_i\). Evidently, as depicted in Fig. \ref{fig:s2f2}, \(F_i\) constitutes a polyhedron. The volume enclosed by the exterior surface of \(F_i\) represents the spatial coverage of the associated facet.

Delaunay tetrahedralization can be robustly executed using the CGAL library \cite{cgal:eb-23b}.

\begin{figure}[htb]
 \begin{minipage}[c]{0.5\textwidth}
    \centering
    \includegraphics[width=2.5 in]{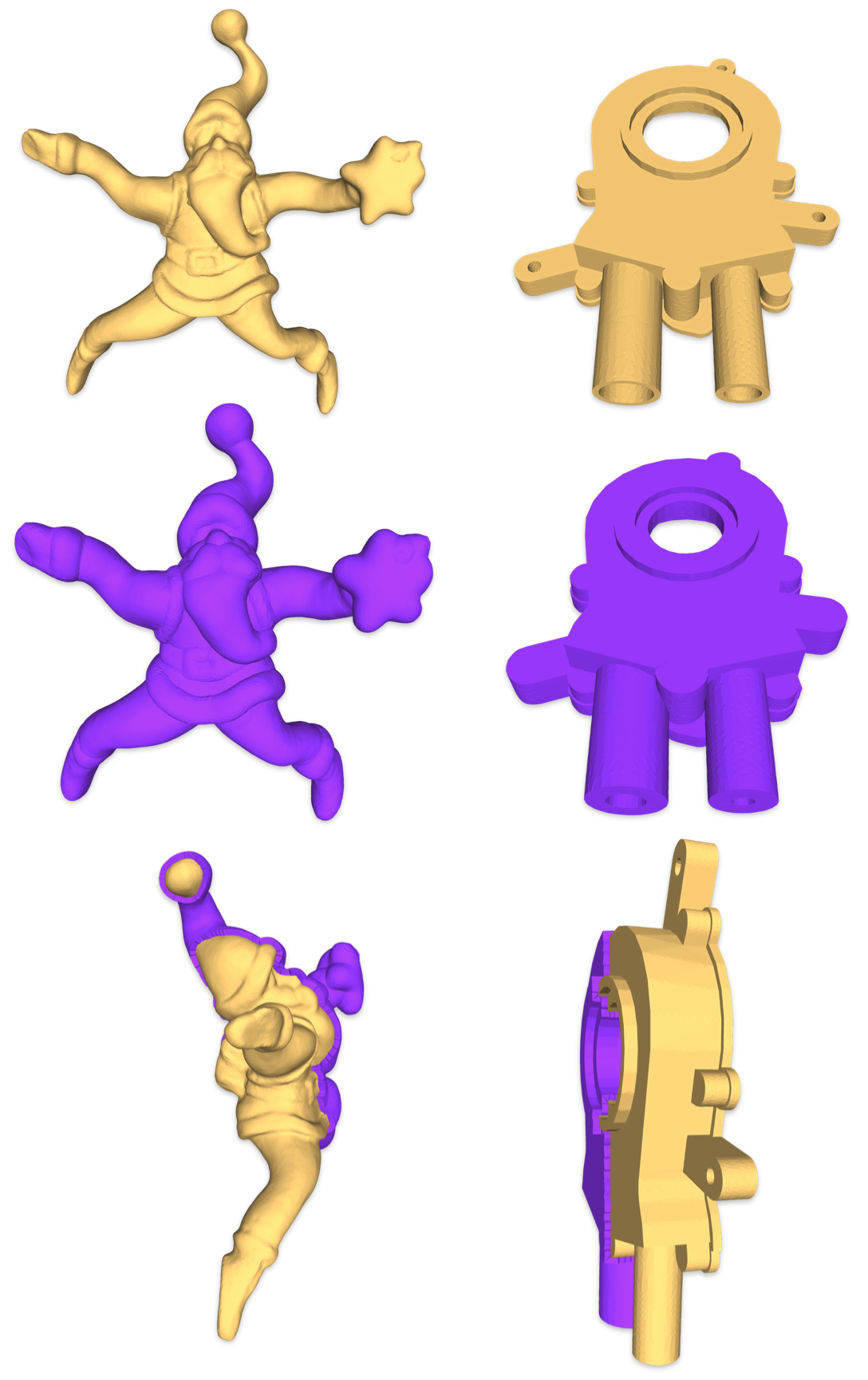}
 \end{minipage}\\

 \caption{The top row illustrates the initial mesh($Mesh_0$). The middle row showcases the cumulative spatial coverage, while the bottom row presents a cross-sectional view of the said cumulative spatial coverage.}\label{fig:32gouzaowan}
\end{figure}

When juxtaposed with the direct offset method, our approach inherently multiplies the number of facets by several magnitudes.
This surge in facet count poses computational challenges, especially during intersection calculations.

Given this, our subsequent discourse seeks to streamline the process of extracting the characteristics of \( Mesh_1 \) from the cumulative spatial coverage. The cumulative spatial coverage is an aggregation of spatial coverage from all facets in \( Mesh_0 \), as illustrated in Fig. \ref{fig:32gouzaowan}.

Upon meticulous examination, we discern that numerous facets can be quickly and definitively judged as non-contributors to the formation of \( Mesh_1 \). This judgment procedure is termed "facet exclusion based on 1-ring neighbor spatial coverage." In what follows, we will utilize the spatial coverage of \(face_i\) as an illustrative example.

\begin{figure}[htb]
 \begin{minipage}[c]{0.5\textwidth}
    \centering
    \includegraphics[width=2.5 in]{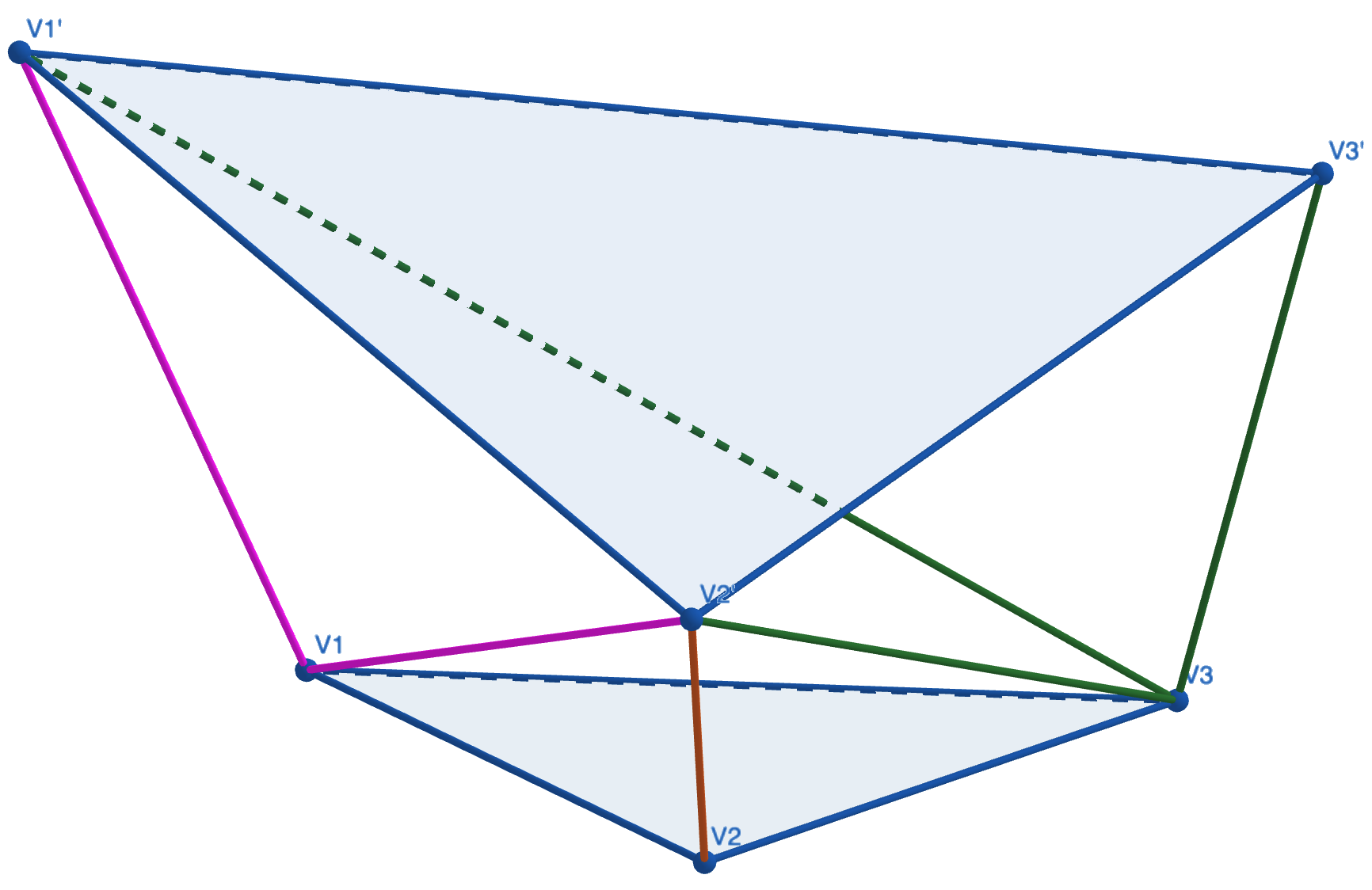}
 \end{minipage}\\

 \caption{In this figure, there are three vertices of the $i$-th facet in $Mesh_0$ are $V_1$, $V_2$ and $V_3$ which calculated offset vertices are $V_1^{'}$, $V_2^{'}$ and $V_3^{'}$.
 The polyhedron representing the spatial coverage is composed of vertices \(V_1\), \(V_2\), \(V_3\), \(V_1'\), \(V_2'\), and \(V_3'\).}\label{fig:s2f1}
\end{figure}

This spatial coverage encompasses six vertices: \(V_1\), \(V_2\), \(V_3\), \(V_1'\), \(V_2'\), and \(V_3'\), as shown in Fig. \ref{fig:s2f1}. Naturally, this foundational logic remains consistent, even for spatial coverage with more vertices. Since the facet \( V_1V_2V_3 \) belongs to the $Mesh_0$, obviously, it cannot serve as a constituent of \( Mesh_1 \).
Consider other facets such as \( V_1V_2V_2' \) and \( V_1'V_2'V_3' \). They may encounter a situation in which they are located on the boundary or within the spatial coverage of \(face_j\), provided that \(face_j\) is adjacent to \(face_i\). If such a situation does occur, they are conclusively deemed impossible as components for $Mesh_1$.

\begin{figure}[htb]
 \begin{minipage}[c]{0.5\textwidth}
    \centering
    \includegraphics[width=2.2 in]{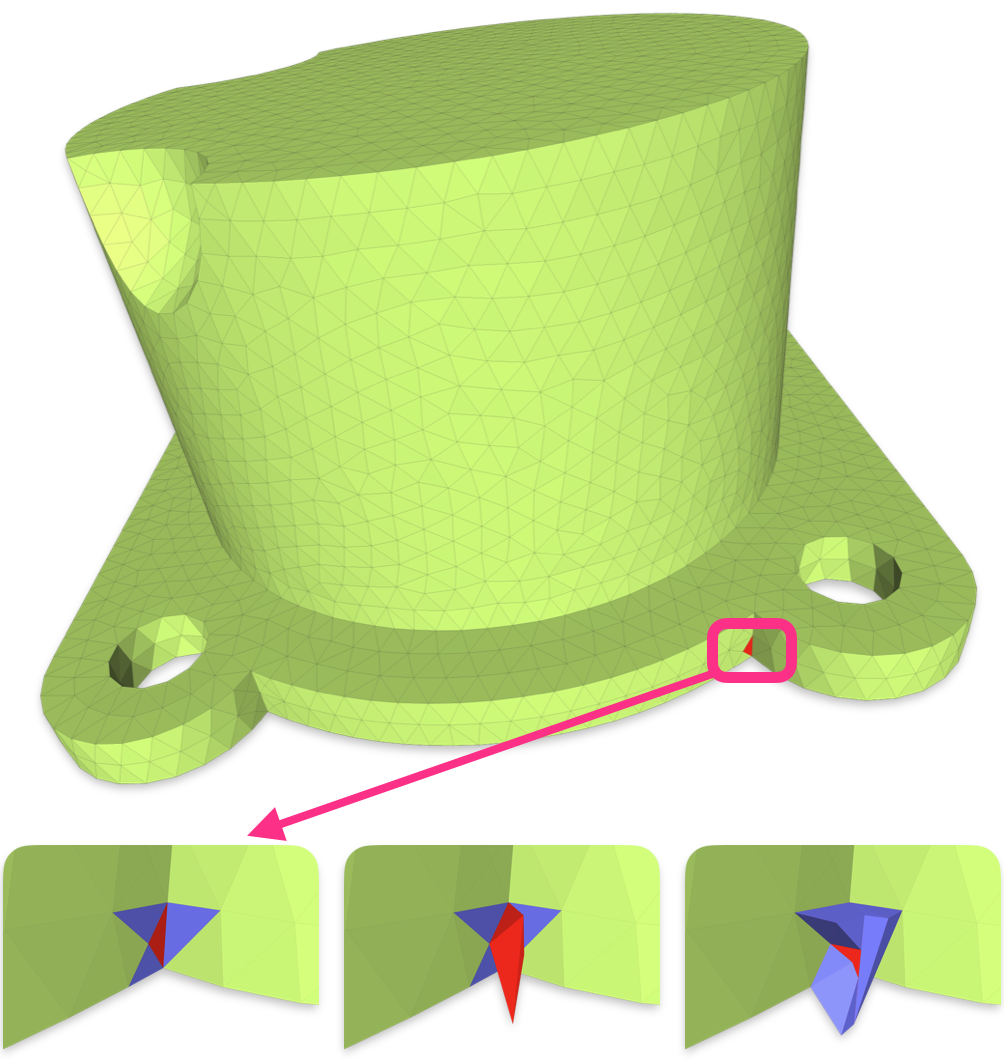}
 \end{minipage}\\

 \caption{This figure illustrates the process of facet exclusion based on 1-ring neighbor spatial coverage. In the bottom-left image, the red triangle represents \(face_i\), while the blue triangles represent \(face_j\). The image in the bottom center depicts \(F_i\), represented as a red polyhedron. The bottom-right image showcases \(F_j\), represented as a blue polyhedron. After eliminating the facets of the red polyhedron that are covered by the blue polyhedron, it becomes evident that only one facet of the red polyhedron remains not be excluded.
}\label{fig:s2ff}
\end{figure}

To further elucidate our specific judgment method as shown in Fig. \ref{fig:s2ff}, we continue to use the facet \( V_1V_2V_2' \) as a representative example. Assuming that \(face_u\), \(face_v\), and \(face_w\) in \( Mesh_0 \) are adjacent to \(face_i\), the next course of action involves calculating non-edge intersection segments between the facet \( V_1V_2V_2' \) and facets from \( F_u \), \( F_v \), and \( F_w \). Utilizing these intersections, facet \( V_1V_2V_2' \) undergoes subdivision into smaller triangular facets via Constrained Delaunay triangulation on a 2D plane.

In scenarios where all subdivided facets are situated within \( F_u \), \( F_v \), or \( F_w \) or lie on their boundaries, the facet \( V_1V_2V_2' \) is confirmed as impossible to be a component of \( Mesh_1 \).

Leveraging this property can significantly reduce computational complexity. Because some triangular facets that are non-contributory can be excluded from the AABB tree before executing the computation of intersections in the Step $S_4$.

\textbf{Note:} Edge intersection means when two triangles intersect, the intersection line coincides with the edge of each triangle.



%% file: 4.4.tex
\subsection{Build Grid and Parallel Computation of Intersections} 
\label{subsec:s4}

\begin{figure}[ht]
 \begin{minipage}[c]{0.5\textwidth}
    \centering
    \includegraphics[width=2.5 in]{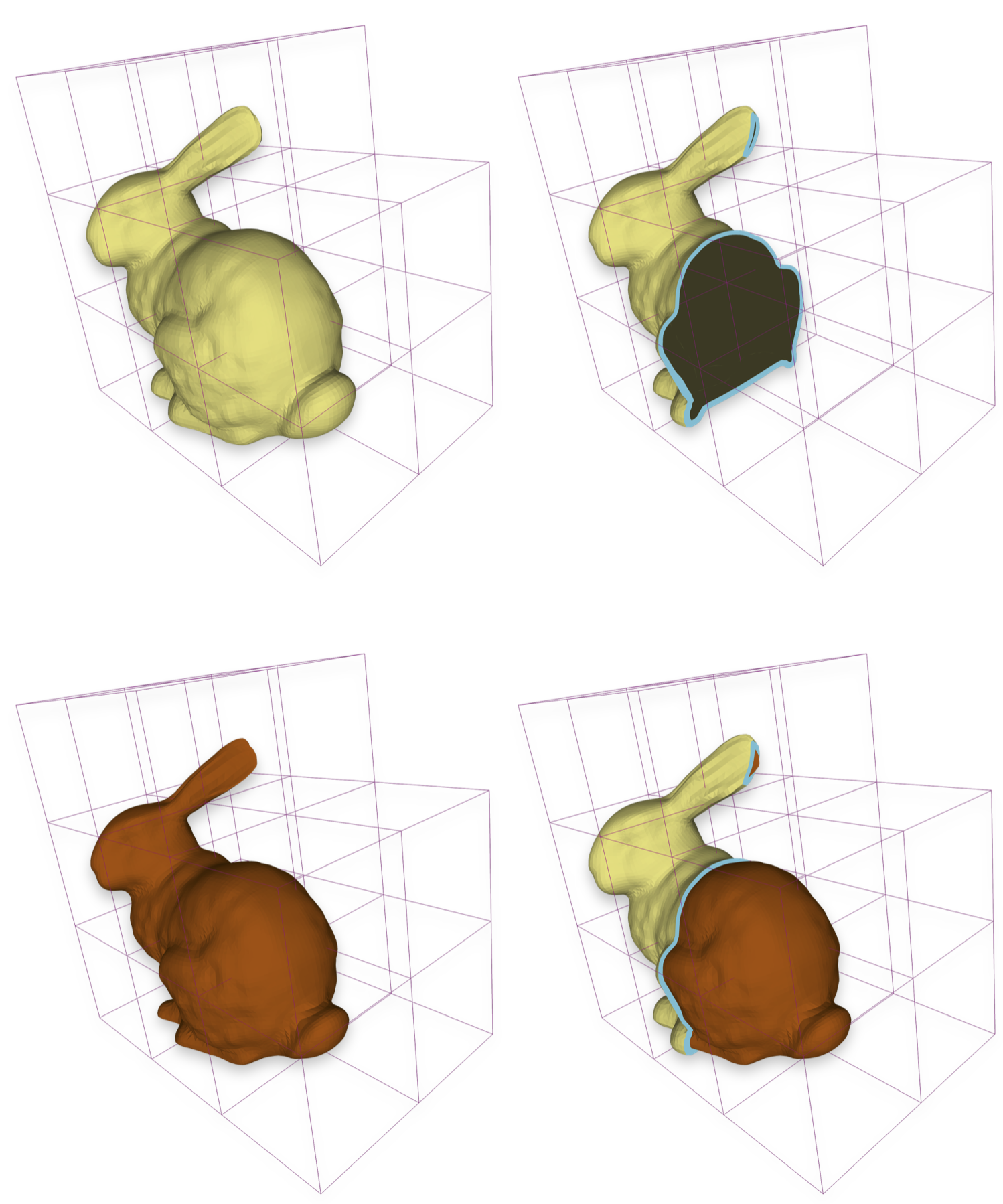}
 \end{minipage}\\

 \caption{From prior steps, the cumulative spatial coverage was derived from \(Mesh_0\). Subsequent procedures aim to extract \(Mesh_1\) from the cumulative spatial coverage. When a mesh that necessitates an inward offset, the figure presented provides a comprehensive visual representation of the relationships among the grid built in this step, \(Mesh_0\), and \(Mesh_1\). The figure is divided into four distinct sections. Top-left: Showcases the grid in conjunction with \(Mesh_0\). Bottom-left: Showcases the grid in conjunction with \(Mesh_1\). Top-right: Features the grid in conjunction with a cross-sectional view, which is the intermediate space between \(Mesh_0\) and \(Mesh_1\). Bottom-right: Combines the cross-sectional view with \(Mesh_1\).
}\label{fig:s3f2}
\end{figure}

In pursuit of heightened efficiency and improved parallel processing, we employ a spatial grid to partition the computational domain into discrete cells, which are geometrically rendered as cubes. This structured decomposition of space facilitates parallel execution by ensuring each cell can be processed independently.

Each cube, characterized by its six faces, has faces that align parallel to the standard three-dimensional planes: \(xy\), \(yz\), and \(zx\).
We designate the vertex with the smallest coordinates in a cell as \(C_0\), and the one with the largest as \(C_1\).
The cell that has the smallest coordinate value for its \(C_0\) vertex is termed \(S_{cell}\). The \(C_0\) vertex coordinates at \(S_{cell}\) represented as \(SC0 = (SC0_x,SC0_y,SC0_z)\).
The coordinates of \(SC0\) are determined by the cumulative spatial coverage.
To elaborate, \(SC0_x\) is the minimum x-coordinate across all vertices in the cumulative spatial coverage, while \(SC0_y\) and \(SC0_z\) are determined similarly.

Given a computational environment with a server equipped with \(c\) cores, our goal is to employ a grid-based spatial decomposition such that the resultant cell count is proximity to \(n \times c\), where \(2 \leq n \leq 40\).
This proposed cell count serves as an aspirational benchmark indicative of optimal performance.
To ensure a more evenly distributed computational load on each core, we assign multiple cells to each core. Therefore, our target is not merely \(c\), but rather \(n \times c\).

The edge length of each cell, is denoted as \(L_g\). Leveraging \(S_{cell}\) as the reference, we position new cells in ascending coordinate direction until the entirety of all spatial coverage is enclosed by the grid as shown in Fig. \ref{fig:s3f2}.
Any given cell can be denoted by \( (n_x, n_y, n_z) \), yielding the vertices \(C_0\) and \(C_1\):

\begin{subequations}
\begin{align}
C_0: (n_x \cdot L_g, n_y \cdot L_g, n_z \cdot L_g) \\
C_1: ((n_x+1) \cdot L_g, ( n_y+1) \cdot L_g, (n_z+1) \cdot L_g)
\end{align}
\end{subequations}


Each cell, denoted generically as \(cell_x\), includes an associated array \(List_x\) that stores handles of certain facets of $Mesh_0$.
The spatial coverage of the facets in \(List_x\) overlap with the space designated by the cell. it's noteworthy that not all cells require instantiation, only those with a non-zero size for \(List_x\) are deemed necessary.

So, we had decomposed the computational domain into several cells.
Each cell houses an AABB tree, denoted as \(AABB_x\) for the cell \(cell_x\).

Every facet of each spatial coverage polyhedron, excluding those eliminated in Step \(S_2\), must be inserted into the AABB tree of its respective cell. If a facet spans multiple cells, it is individually inserted into the AABB trees of these cells.

The subsequent part of this step is the parallel computation of intersections among all facets. Distinct threads execute this for every cell.
Within each cell, facet intersections are expedited via its inherent AABB tree.
All intersections, both points and lines, serve as constraints.
These constraints assist in the Delaunay triangulation of the facets, further subdividing them into smaller facets.

Once every facet undergoes this procedure, all spatial coverage facets are transformed into these subdivided facets.
Relying on these subdivided facets, we reconstruct each spatial coverage.
For a given spatial coverage \(F_i\), its \(j^{th}\) facet is labeled \(S_{ij}\).
Clearly, intersections between \(S_{ij}\) and other subdivided facets will not occur within the facet itself. Namely, when two facets intersect, the line of intersection invariably lies on the boundaries of both facets.

%% file: 4.5.tex
\subsection{Extraction of $Mesh_1$ from Cumulative Spatial Coverage}
\label{subsec:s5}
\begin{figure}[htb]
 \begin{minipage}[c]{0.5\textwidth}
    \centering
    \includegraphics[width=2.8 in]{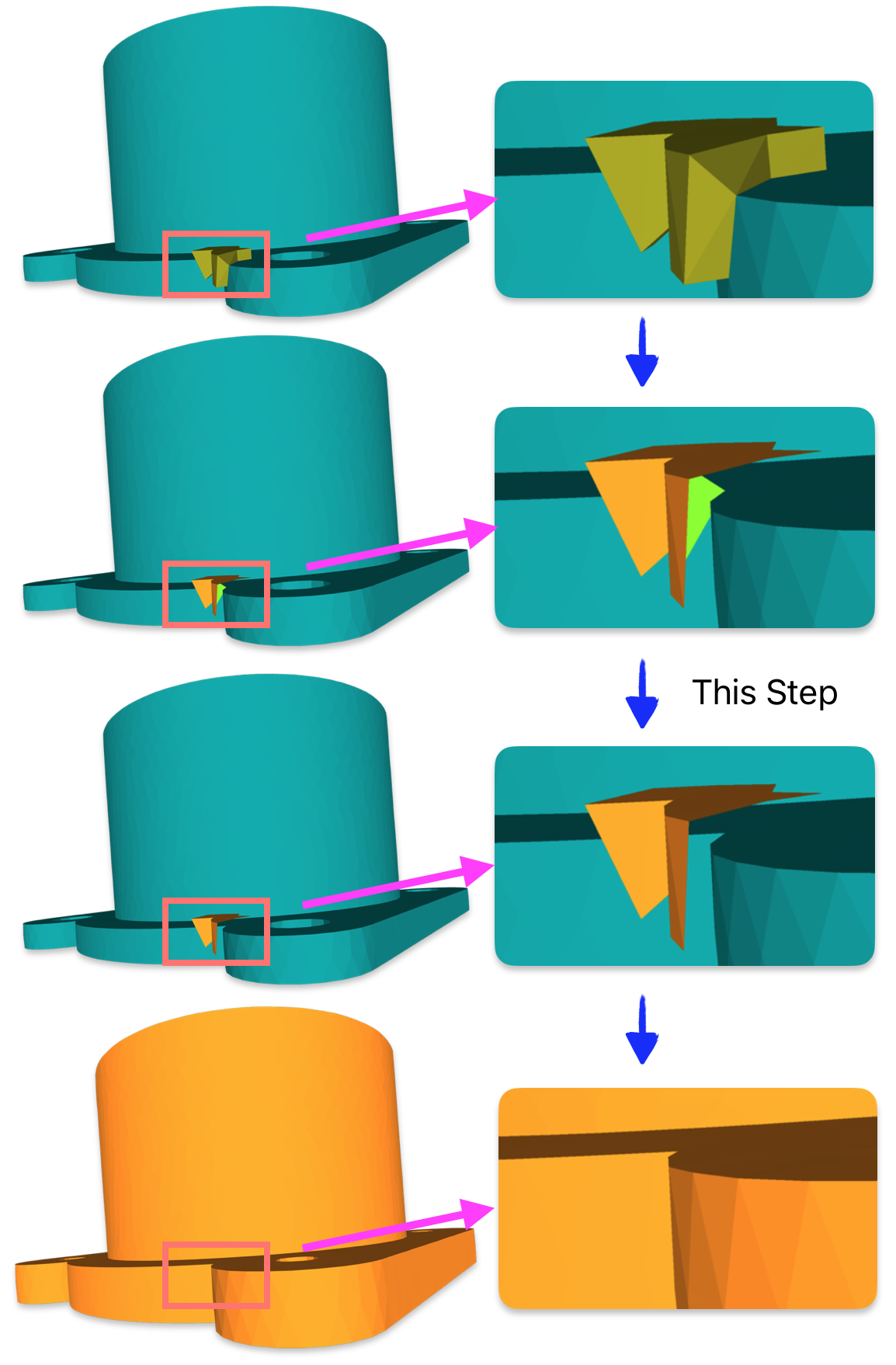}
 \end{minipage}\\
 \caption{
     The presented figure is divided into four distinct sections, proceeding from top to bottom.
     The transition from the first to the second section encapsulates the process delineated as the exclusion of facets via 1-ring neighbor spatial coverage, as detailed in subsection \ref{subsec:s3}.
     The progression from the second to the third section underscores the primary utility of this step. Within the graphic, the green triangular facet denote those that couldn't be excluded in subsection \ref{subsec:s3}.
     However, through the employment of an AABB tree and ray detection, these facets are effectively excluded during this step.
     The transition from the third to the fourth section illustrates the consolidation of all localized facets, culminating in the resultant offset surface represented as an unordered collection of triangular facets, or Triangle Soup.}\label{fig:s5f1}
\end{figure}
In this subsection, our goal is to extract $Mesh_1$ from the cumulative spatial coverage as shown in Fig. \ref{fig:s5f1}.
This procedure, which is inherently parallelizable, operates independently for each cell and is divided into two critical stages.

\noindent\textbf{Stage One: Ray-Based detection}

Within the spatial domain of each cell, numerous instances of spatial coverage are found. For each instance, its facets, one of which is denoted as \(S_{ij}\), are inserted into the cell's axis-aligned bounding box (AABB) tree. The AABB tree corresponding to the \(x^{th}\) cell is represented as \(AABB2_x\).
Distinct from the AABB tree utilized in Step $S_4$, referenced in subsection \ref{subsec:s4}, \(S_{ij}\) is inserted into the AABB tree even if not contained within the spatial domain of the cell, as long as \(F_i\) traverses through or contained by the cell's space.

 \(S_{ij}\) is derived by the subdivision of a specific facet in Step $S_4$, as referenced in subsection \ref{subsec:s4}, it follows that if the specific facet is excluded during the process delineated in Step $S_2$ (subsection \ref{subsec:s2}), then \(S_{ij}\) can be definitively ruled out as a potential constituent of \(Mesh_1\).

For the facet \(S_{ij}\) that have not been excluded, a subsequent evaluation is required:

If the facet is situated within a single instance of spatial coverage, or if it aligns with the boundaries of two instances of distinct spatial coverage, with each coverage distinctly positioned on opposite sides of \(S_{ij}\), then such a facet cannot be a component of \(Mesh_1\).

To streamline and expedite this evaluative process, we introduce a ray-casting methodology.
For facets \(S_{ij}\) located in the spatial domain of \(cell_x\), a ray is emanated from the geometric centroid of \(S_{ij}\), cast in a non-prescriptive direction to do fast intersection check with all facets saved in \(AABB2_x\).
Then, Based on the intersections computed above, we can determine whether \(S_{ij}\) serves as a component of \(Mesh_1\). If \(S_{ij}\) is present in multiple cells, the intersections derived from these cells must be collectively considered to make a determination.






\noindent\textbf{Stage Two: check each $S_{ij}$ internal/external in the $Mesh_0$}

Assuming an inward offsetting approach is taken. During extensive offsetting operations, certain facets might emerge outside of \(Mesh_0\).
Clearly, these facets are not constituent elements of \(Mesh_1\) and thus necessitate exclusion.
For \(Mesh_0\) that is both closed and manifold, the CGAL \cite{cgal:eb-23b} library offers a robust mechanism to facilitate this determination. If not, using winding number\cite{jacobson2013robust} to realize internal and external judgment.

If we do the outward offset, just change the check if internal to check if external.

\noindent\textbf{Post processing}

\begin{figure}[htb]
 \begin{minipage}[c]{0.5\textwidth}
    \centering
    \includegraphics[width=2.8 in]{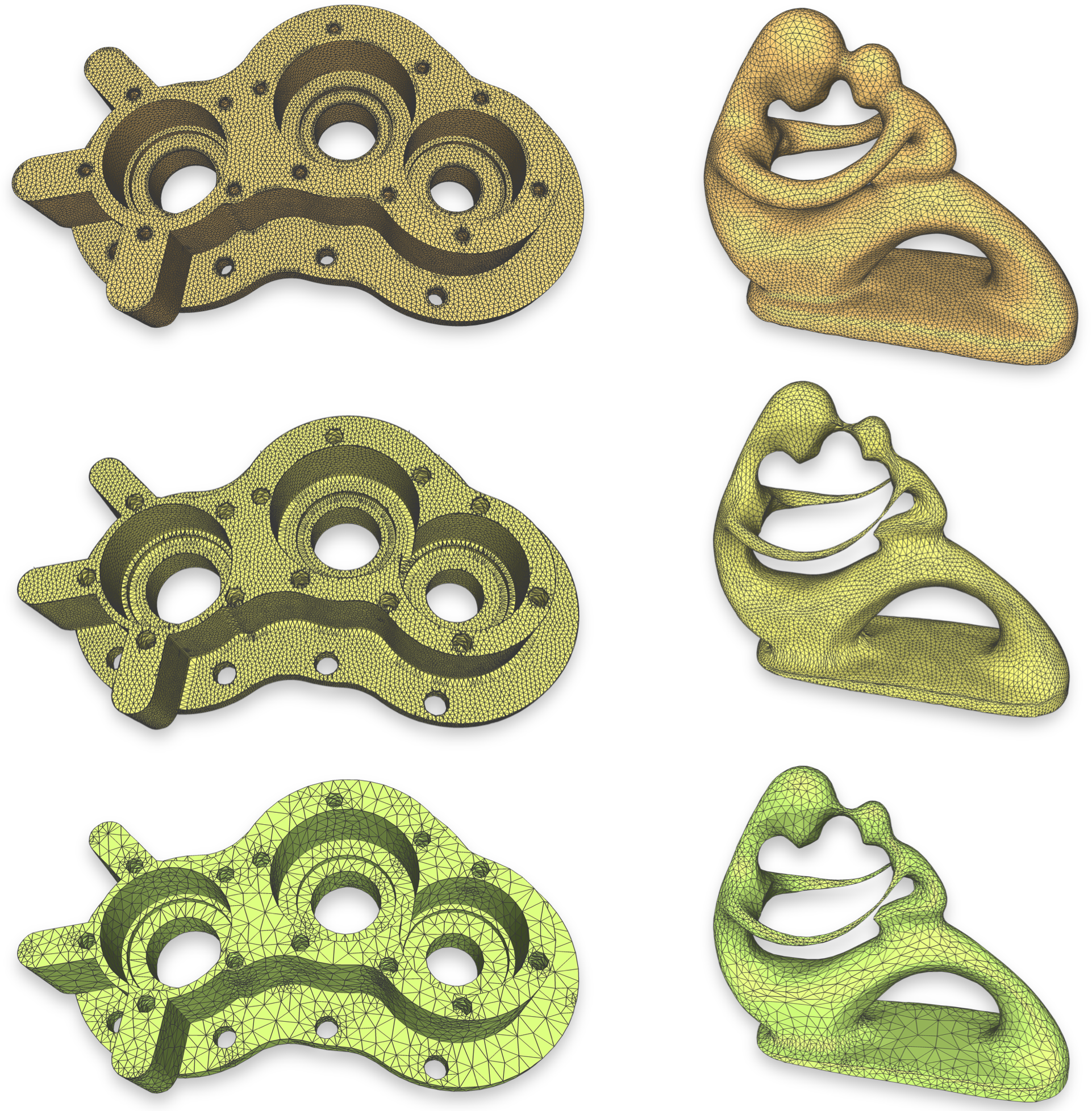}
 \end{minipage}\\
 \caption{The figure elucidates the results of an inward offset.
 The top row delineates the initial mesh($Mesh_0$), the middle row showcases the resultant Triangle Soup, and the bottom row conveys the final offset mesh($Mesh_1$), as generated by TetWild.}\label{fig:s5octin}
\end{figure}

\begin{figure}[htb]
 \begin{minipage}[c]{0.5\textwidth}
    \centering
    \includegraphics[width=2.8 in]{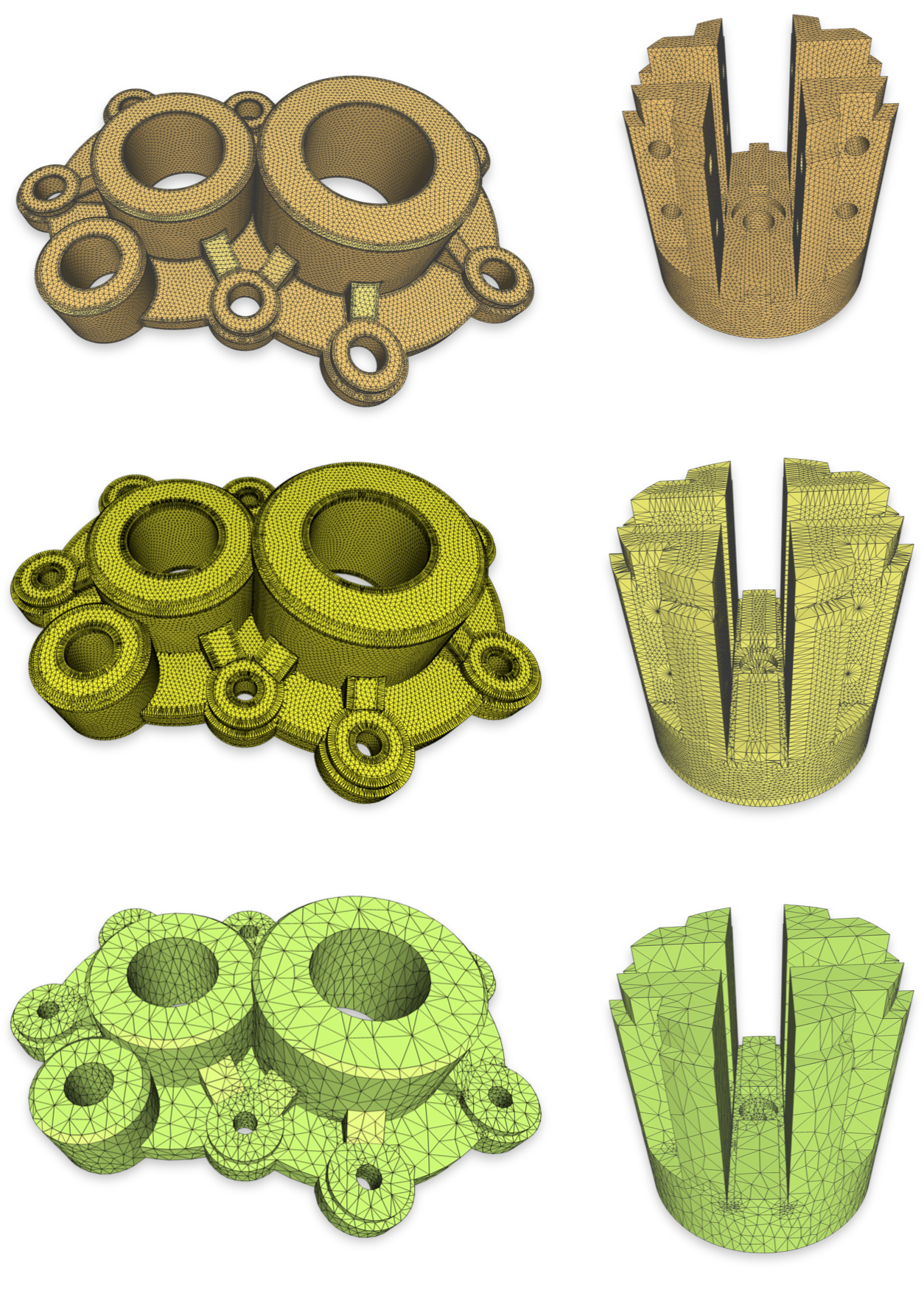}
 \end{minipage}\\
 \caption{The figure elucidates the results of an outward offset.
 The top row delineates the initial mesh($Mesh_0$), the middle row showcases the resultant Triangle Soup, and the bottom row conveys the final offset mesh($Mesh_1$), as generated by TetWild.
}\label{fig:s5octout}
\end{figure}

Upon completion of the aforementioned stages, a Triangle Soup is generated.
Finally, show as Fig. \ref{fig:s5octin} and Fig. \ref{fig:s5octout}, TetWild \cite{hu2018tetrahedral} are used to build mesh.

If the quality of the generated mesh are dissatisfied, it is recommended to carry out a final remeshing procedure. Several remeshing strategies are available, as highlighted in \cite{nivoliers2015anisotropic}, \cite{kobbelt2004remeshing}, \cite{hu2016error}, \cite{alliez2003isotropic}, \cite{wang2018isotropic}, \cite{botsch2004remeshing}, and \cite{dunyach2013adaptive}.
For instance, the methods presented in \cite{botsch2004remeshing} and \cite{dunyach2013adaptive} have been robustly implemented in the pmp-library \cite{pmp-library}.

%% file: ExperimentalResults.tex
 \section{Experimental results}
 \label{sec:experiment}

\begin{figure}[htb]
 \begin{minipage}[c]{0.5\textwidth}
    \centering
    \includegraphics[width=2.5 in]{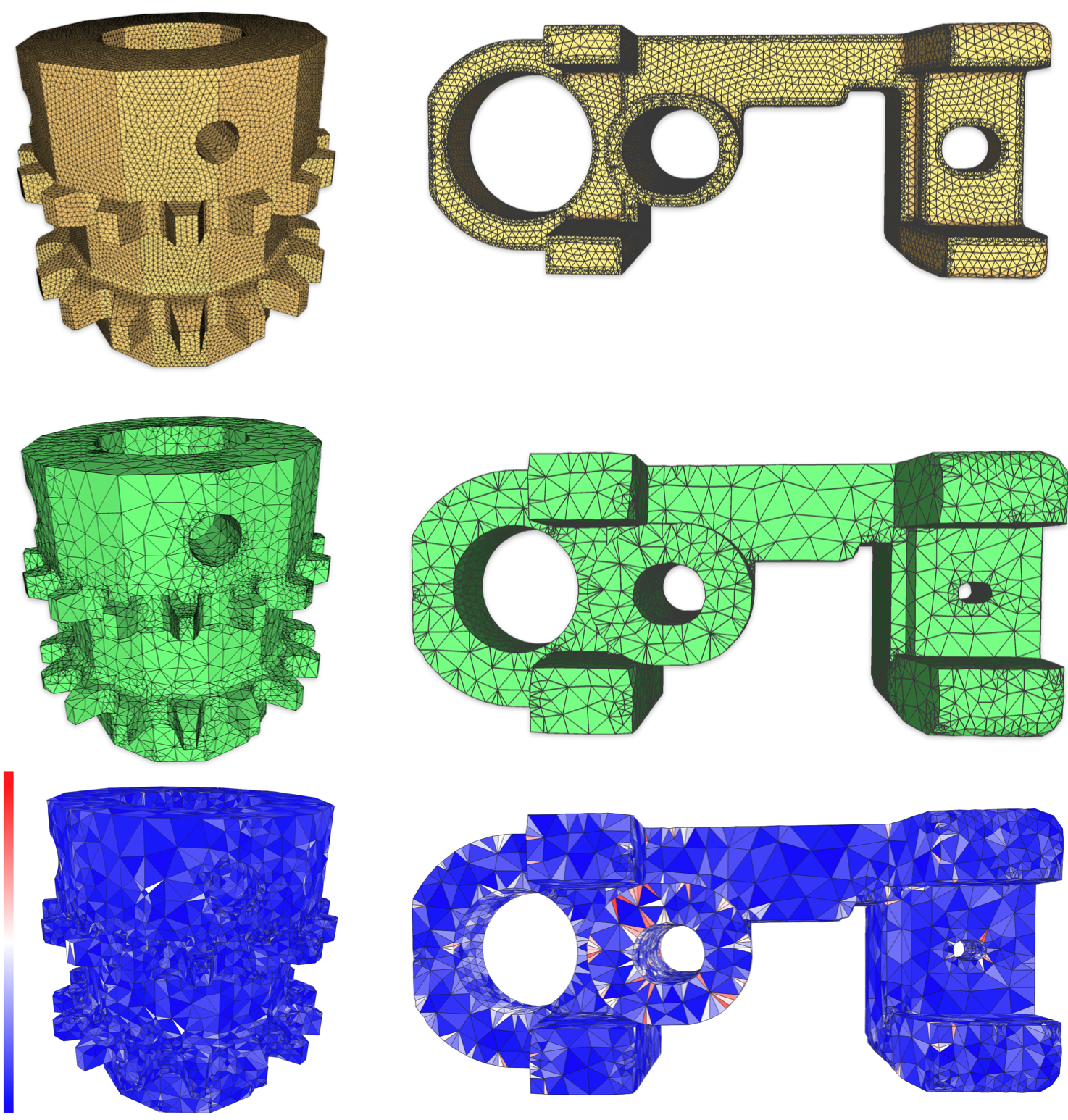}
 \end{minipage}\\
 \caption{
The figure is structured as follows: The topmost row portrays the initial mesh, denoted as \(Mesh_0\).
     Subsequently, the intermediate row presents the concluding offset mesh, represented as \(Mesh_1\).
     The final row serves as an illustrative representation of mesh quality. Facets shaded in a blue hue signify superior quality, indicating a favorable aspect ratio, while those leaning towards a red coloration represent quality with a less desirable aspect ratio.
     It is noteworthy that a significant proportion of the triangular facets generated via TetWild demonstrate high-quality characteristics.}\label{fig:s5f3}
\end{figure}

\begin{figure}[htb]
 \begin{minipage}[c]{0.5\textwidth}
    \centering
    \includegraphics[width=2.5 in]{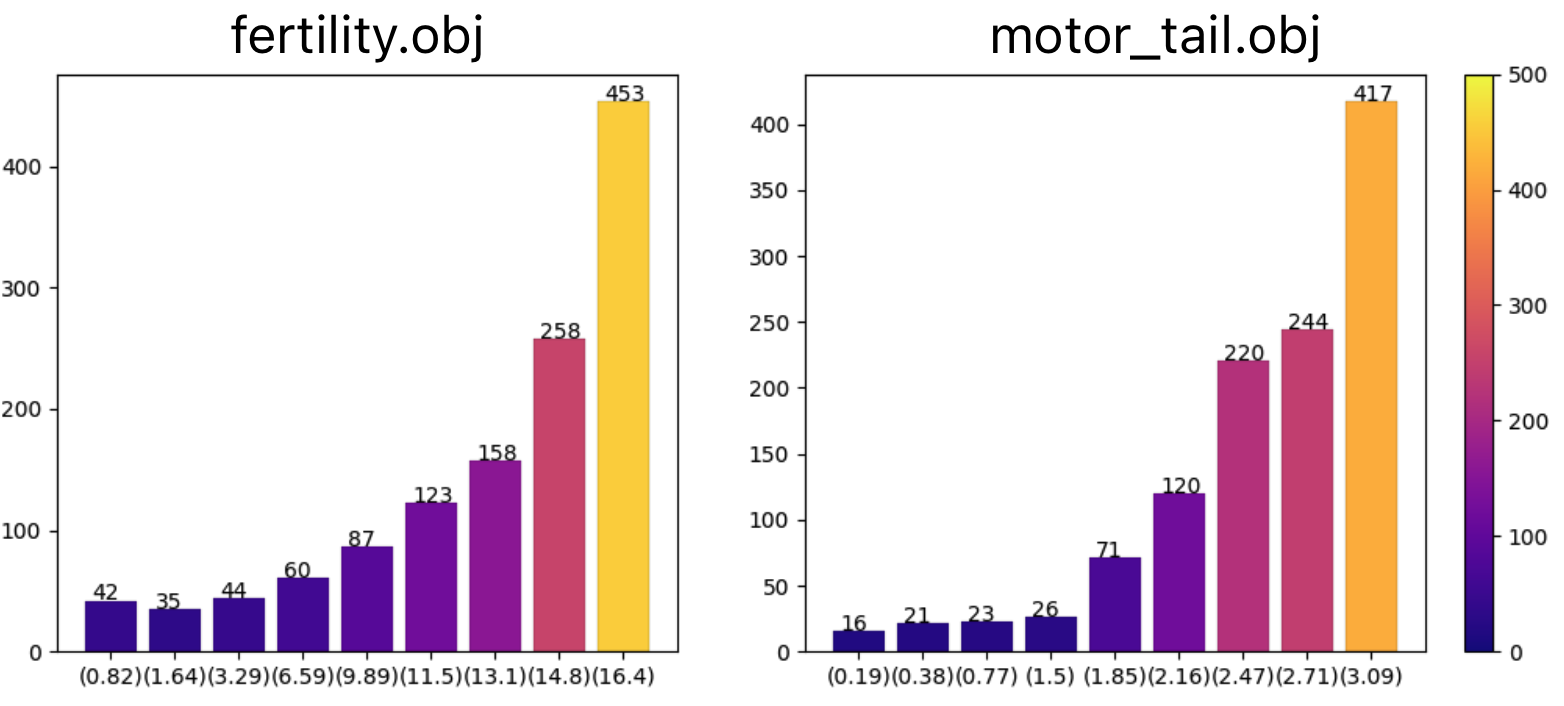}
 \end{minipage}\\
 \caption{
For simplicity, we used an invariable mesh offset as an example. For each model, we conducted multiple tests with varying offset distances to measure the time required.
The horizontal axis denotes the offset distance, while the vertical axis represents time, measured in seconds.
     It is discernible that as the offset distance incrementally increases, the efficiency of our algorithm markedly diminishes.
     This decrease in performance can be fundamentally attributed to the intricate triangular facet intersections that arise with extended offset distances.
 }\label{fig:octspeed}
\end{figure}

 We use the C++ programming language with GNU C++ compiler to implement the presented algorithm. The test device has 64GB memory, with the CPU of Intel i9-10900K, the OS of it is Ubuntu 20.04.

\begin{table*}
    \centering
    \begin{tabular}{lrrrr|rrr|rrr}
        \toprule
             & & & & &\multicolumn{3}{c}{Offset outward} & \multicolumn{3}{|c}{Offset inward} \\
        \textbf{model name} & \textbf{vertex} & \textbf{face} & \textbf{$D_{min}$} & \textbf{$D_{max}$} & \textbf{vertex} & \textbf{face} & \textbf{time} & \textbf{vertex} & \textbf{face} & \textbf{time} \\
        \midrule
        BendTube & 2843 & 5690 & 5.27888 & 10.55780 & 2001 & 4006 & 11.0 & 1854 & 3704 & 22.4 \\
        CoverSlipCleaningHolder2b & 12653 & 25506 & 0.65915 & 1.31830 & 4081 & 8362 & 566.2 & 5397 & 10794 & 3079.3 \\
        Hub & 6450 & 12980 & 1.65474 & 3.30948 & 5150 & 10380 & 108.9 & 2590 & 5188 & 78.7 \\
        LowerControlArm & 9317 & 18642 & 1.35551 & 2.71103 & 4172 & 8352 & 30.4 & 4309 & 8624 & 226.7 \\
        anti\_backlash\_nut & 6388 & 12784 & 0.36751 & 0.73501 & 2253 & 4514 & 38.7 & 2671 & 5350 & 83.1 \\
        asm007 & 10019 & 20042 & 0.00241 & 0.00482 & 2978 & 5960 & 82.2 & 3527 & 7058 & 550.0 \\
        bamboo\_pen & 17832 & 35692 & 0.30289 & 0.60577 & 3549 & 7126 & 74.6 & 3720 & 7468 & 231.8 \\
        bearing\_plate & 4907 & 9830 & 0.53945 & 1.07889 & 1435 & 2886 & 22.3 & 2092 & 4200 & 101.4 \\
        beveled\_shoulder\_1 & 3836 & 7668 & 0.57826 & 1.15651 & 415 & 826 & 14.9 & 356 & 708 & 223.6 \\
        beveled\_shoulder\_2 & 4106 & 8212 & 0.59610 & 1.19220 & 461 & 922 & 24.6 & 459 & 918 & 232.7 \\
        blade & 14532 & 29060 & 0.00511 & 0.01023 & 3227 & 6450 & 48.0 & 4111 & 8218 & 42.8 \\
        bladefem & 3621 & 7238 & 0.35489 & 0.70977 & 1235 & 2466 & 26.0 & 1039 & 2074 & 42.5 \\
        bolt & 4752 & 9504 & 0.11209 & 0.22418 & 1370 & 2740 & 16.6 & 1720 & 3440 & 16.8 \\
        bone & 4666 & 9328 & 0.00643 & 0.01287 & 1925 & 3846 & 16.4 & 1934 & 3864 & 15.4 \\
        bone1 & 1566 & 3128 & 0.01205 & 0.02409 & 1249 & 2494 & 6.1 & 1428 & 2852 & 6.0 \\
        bozbezbozzel50K & 24992 & 50000 & 1.39455 & 2.78910 & 11206 & 22418 & 1529.5 & 12755 & 25506 & 591.8 \\
        buddha & 63264 & 126524 & 0.00444 & 0.00887 & 14247 & 28490 & 1010.4 & 15919 & 31834 & 220.7 \\
        bunny & 11120 & 22236 & 0.00521 & 0.01042 & 4485 & 8966 & 56.9 & 4498 & 8992 & 36.1 \\
        casting & 9484 & 19000 & 0.00839 & 0.01677 & 4485 & 9002 & 56.1 & 6192 & 12466 & 157.4 \\
        casting2 & 9513 & 19058 & 0.00837 & 0.01674 & 4251 & 8534 & 78.6 & 6452 & 12992 & 157.0 \\
        cow2 & 4315 & 8626 & 0.01350 & 0.02701 & 3769 & 7534 & 65.8 & 3696 & 7388 & 121.5 \\
        cube\_minus\_sphere & 2070 & 4136 & 0.01555 & 0.03111 & 1017 & 2030 & 6.0 & 1064 & 2124 & 24.9 \\
        cup-tri & 5668 & 11340 & 0.38868 & 0.77737 & 3148 & 6300 & 33.0 & 3551 & 7106 & 26.7 \\
        deckel & 5024 & 10060 & 0.38905 & 0.77809 & 2315 & 4642 & 17.7 & 2079 & 4174 & 23.1 \\
        delta\_arm2\_remesh & 48668 & 97348 & 0.20622 & 0.41244 & 1545 & 3102 & 208.4 & 4439 & 8890 & 291.8 \\
        des6 & 3316 & 6632 & 0.87638 & 1.75275 & 1421 & 2842 & 28.3 & 1258 & 2516 & 56.7 \\
        des7 & 11061 & 22126 & 0.26276 & 0.52551 & 1571 & 3148 & 55.7 & 2186 & 4414 & 186.6 \\
        eros100K & 50002 & 100000 & 0.62686 & 1.25373 & 11079 & 22154 & 227.5 & 10830 & 21656 & 189.7 \\
        fandisk & 7250 & 14496 & 0.04936 & 0.09872 & 923 & 1842 & 28.3 & 1044 & 2084 & 81.1 \\
        greek\_sculpture & 24994 & 50000 & 0.76476 & 1.52952 & 11683 & 23378 & 7021.3 & 10852 & 21716 & 3779.0 \\
        gyroidpuzzle & 13121 & 26298 & 0.35096 & 0.70191 & 8052 & 16160 & 38.1 & 8052 & 16160 & 170.3 \\
        halved\_oblique\_scarf\_2 & 5566 & 11128 & 0.45402 & 0.90805 & 565 & 1126 & 24.6 & 462 & 920 & 104.3 \\
        halved\_oblique\_scarf\_3 & 6164 & 12324 & 0.42987 & 0.85973 & 584 & 1164 & 34.8 & 631 & 1258 & 120.4 \\
        hinge & 9544 & 19104 & 0.36614 & 0.73228 & 2341 & 4698 & 30.7 & 2485 & 4986 & 55.3 \\
        impeller & 13063 & 26126 & 0.01834 & 0.03668 & 2795 & 5590 & 55.7 & 3525 & 7050 & 590.0 \\
        inlay\_dovetail\_3 & 7541 & 15078 & 0.62424 & 1.24847 & 759 & 1514 & 30.6 & 1367 & 2804 & 226.9 \\
        lego & 10602 & 21200 & 0.56211 & 1.12422 & 2983 & 5962 & 51.2 & 2038 & 4072 & 267.3 \\
        lion & 17198 & 34392 & 0.00093 & 0.00187 & 7340 & 14676 & 49.9 & 6838 & 13672 & 52.7 \\
        lock\_Lp & 5788 & 11584 & 1.04538 & 2.09076 & 2092 & 4192 & 24.2 & 3429 & 6882 & 252.1 \\
        magalie\_hand100K & 50002 & 100000 & 0.48509 & 0.97018 & 7600 & 15204 & 208.7 & 7589 & 15174 & 168.4 \\
        mech\_piece & 2089 & 4174 & 0.55146 & 1.10292 & 1062 & 2120 & 6.8 & 1017 & 2030 & 66.1 \\
        mid2Fem & 8868 & 17752 & 0.77463 & 1.54925 & 2001 & 4018 & 24.6 & 1970 & 3956 & 94.0 \\
        motor\_tail & 7567 & 15182 & 0.51505 & 1.03010 & 2761 & 5570 & 30.9 & 3096 & 6216 & 85.6 \\
        nasty\_cheese & 30816 & 62160 & 0.44832 & 0.89665 & 23541 & 47614 & 142.0 & 23749 & 47950 & 679.0 \\
        nugear & 17331 & 34658 & 0.21120 & 0.42239 & 2161 & 4318 & 78.5 & 2266 & 4528 & 130.3 \\
        octa\_flower\_Lp & 15143 & 30282 & 0.14846 & 0.29692 & 1712 & 3420 & 66.1 & 2038 & 4072 & 632.3 \\
        part2 & 5341 & 10710 & 4984.49 & 9968.99 & 2968 & 5964 & 38.1 & 3319 & 6670 & 105.3 \\
        part20k & 13055 & 26118 & 0.02221 & 0.04442 & 1866 & 3740 & 114.5 & 2765 & 5538 & 284.5 \\
        pinion & 21439 & 42878 & 0.00925 & 0.01851 & 1613 & 3226 & 54.4 & 2336 & 4672 & 105.3 \\
        \bottomrule
    \end{tabular}
    \caption{The test data was obtained from the website [https://www.quadmesh.cloud]\cite{pietroni2021reliable}. We conducted a variable offset operation where the offset distance uniformly ranged between \(D_{min}\) and \(D_{max}\).
    The time presented in the table signifies the computational expense, and it is measured in seconds.}
    \label{tab:xxx}
\end{table*}

Using CGAL's Exact\_predicates\_exact\_constructions\_kernel to perform the intersection operation to avoid the precision problem of floating-point arithmetic.

The result data is presented in Table \ref{tab:xxx}.
 This data is the execution of a variable offset operation where the offset distance ranges between \(D_{min}\) and \(D_{max}\).
 The offset distance for each face is determined by the x-axis coordinate of its geometric center.
 Let the x-axis coordinates on both sides of the bounding box of \(Mesh_0\) be \(X_{min}\) and \(X_{max}\) respectively.
If the x-axis coordinate of a face is denoted as $X$, then its offset distance can be calculated as:
$ D_{min} + \frac{X - X_{min}}{X_{max} - X_{min}} (D_{max} - D_{min})$.

\begin{figure}[htb]
 \begin{minipage}[c]{0.5\textwidth}
    \centering
    \includegraphics[width=2.7 in]{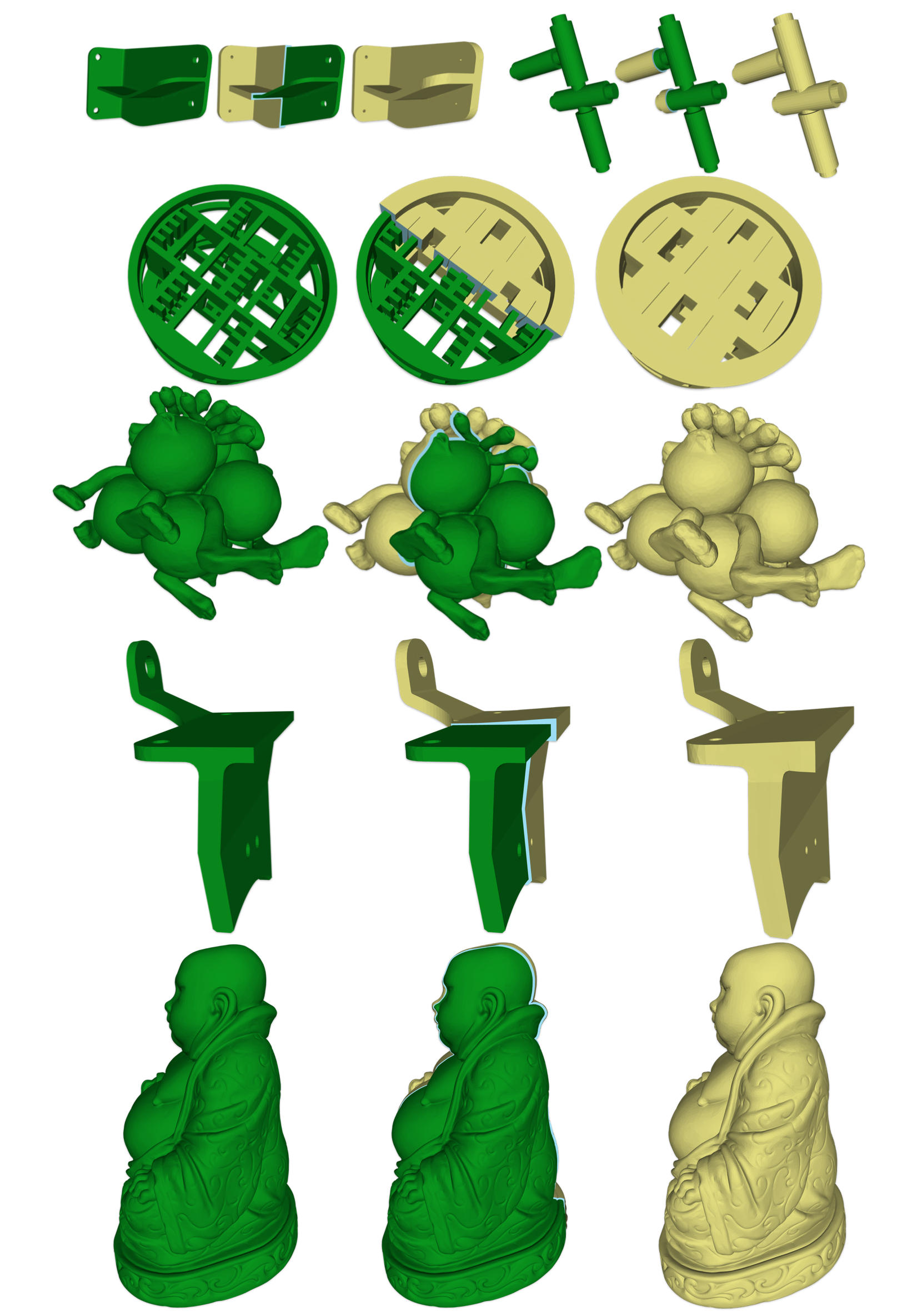}
 \end{minipage}\\
 \caption{The variable offset outward,
  there are six groups in total, and the first mesh in each group is the green mesh, representing the initial triangular mesh.
 The second is a cross-sectional view of the resulting shell-like structure, and the third shows the outer surface.
 }\label{fig:d4f1_3}
\end{figure}

\begin{figure*}[htb]
\begin{minipage}[c]{1.0\textwidth}
    \centering
    \includegraphics[width=5.1in]{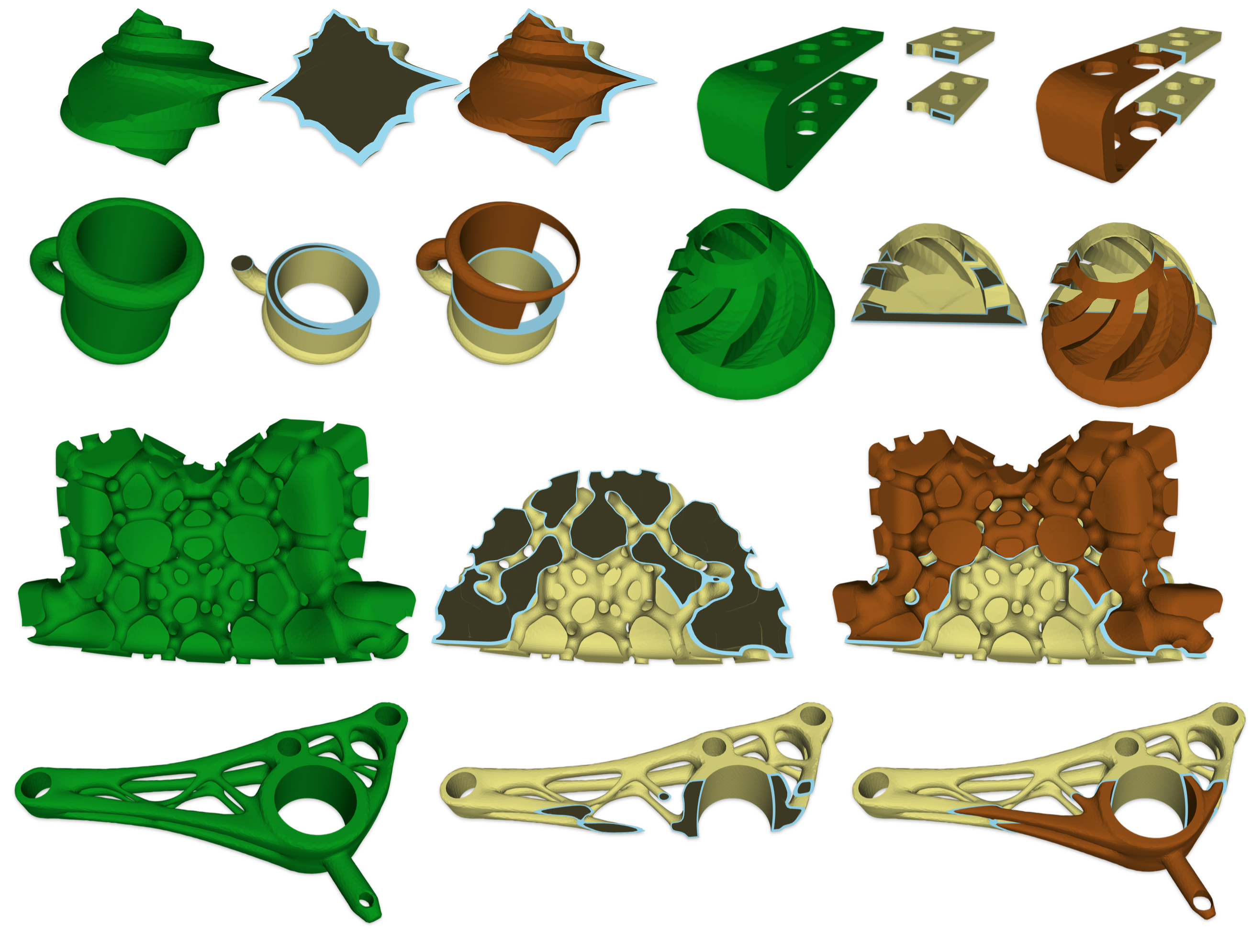}
 \end{minipage}%
 \caption{The variable offset inward, there are six groups in total, and the first mesh in each group is the green mesh, representing the initial triangular mesh.
 The second is a cross-sectional view of the resulting shell-like structure, and the third shows the inner surface and the cross-sectional view together.}\label{fig:d4f1_2}
\end{figure*}

\begin{figure*}[htb]
\begin{minipage}[c]{1.0\textwidth}
    \centering
    \includegraphics[width=5.5in]{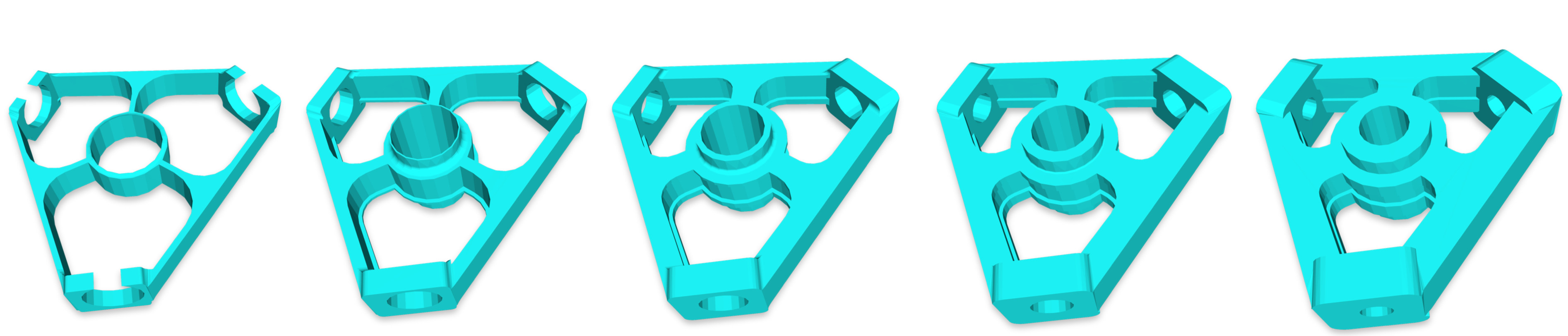}
 \end{minipage}%
 \caption{The middle one is the initial mesh, and the two pictures on the left are the results of the long-distance inward offset and the short-distance inward offset obtained by our algorithm respectively.
  The two figures on the right show the results of short-distance outward offset and long-distance outward offset obtained by our algorithm respectively.}\label{fig:d4f1}
\end{figure*}

\begin{figure*}[htb]
\begin{minipage}[c]{1.0\textwidth}
    \centering
    \includegraphics[width=4 in]{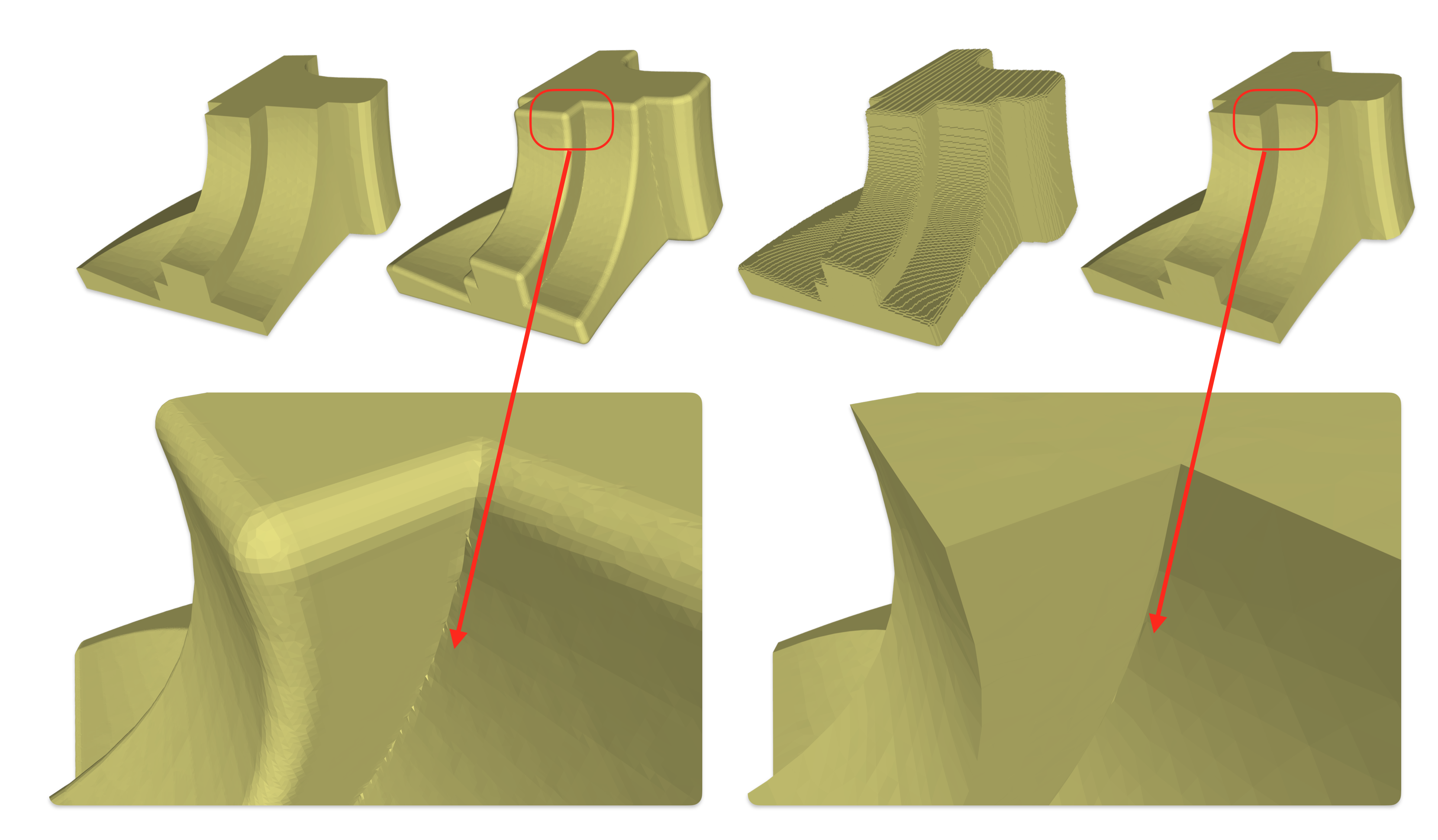}
 \end{minipage}%
 \caption{The first column is the initial mesh, the second column is the outward offset based on the distance field, the third column is the result generated based on rays and dexels, and the fourth column is our method., we can find that the method based on sampling will be distorted due to ambiguity.}\label{fig:d4f2}
\end{figure*}

\begin{figure*}[htb]
\begin{minipage}[c]{1.0\textwidth}
    \centering
    \includegraphics[width=5.2in]{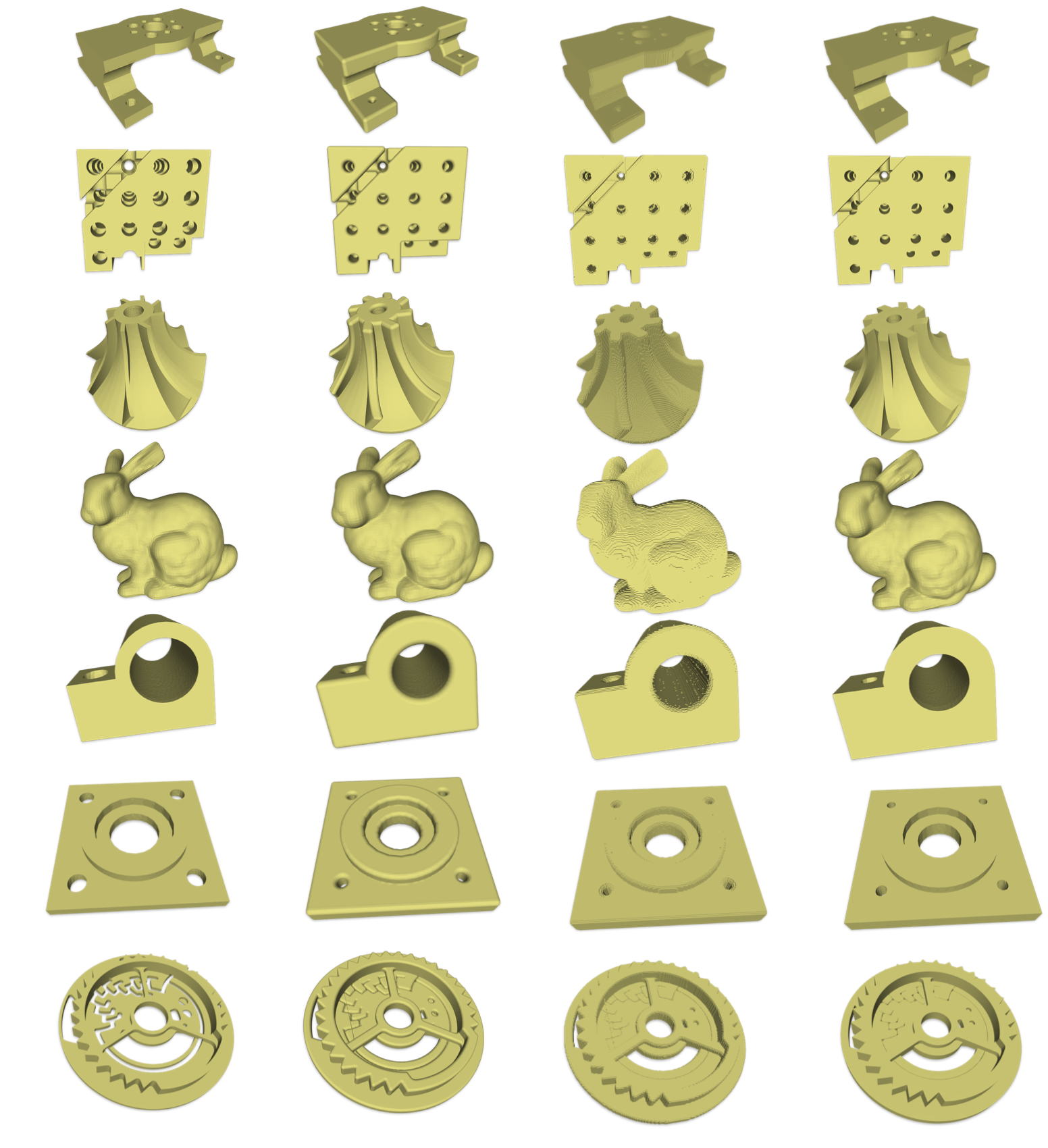}
 \end{minipage}%

 \caption{ The meaning of four column is the same as the figure\ref{fig:d4f2}.
 From the figure, a notable distinction between the result obtained using our method and those derived from existing techniques is the preservation of sharp features.
 Our approach adeptly replicates right-angle edges and vertices.}\label{fig:d4f4}
\end{figure*}

Fig. \ref{fig:s5f3} illustrates the mesh quality, while Fig. \ref{fig:octspeed} displays the runtime of invariable mesh offset with varying offset distances.
Fig. \ref{fig:d4f1_3}, Fig. \ref{fig:d4f1_2}, Fig. \ref{fig:d4f1}, Fig. \ref{fig:d4f2}, and Fig. \ref{fig:d4f4} provide visual representations corresponding to results from our experiments.

From Fig. \ref{fig:d4f1_3}, it can be observed that our method, when applying a variable outward offset to the mesh, retains sharp features.
 Furthermore, when executing short-distance offsets, the mesh remains intact without fragmentation.

Similarly, Fig. \ref{fig:d4f1_2} indicates that when our method applies a variable inward offset to the mesh, it can preserve sharp features without causing any breakage in the mesh.


Currently, our demonstration software does not incorporate procedures for handling instances where \(Mesh_0\) is not closed or not manifold.
 Consequently, for such meshes, it is imperative to first apply corrective algorithms before executing our code.
 Potential mesh-repair methodologies can be found in the following studies: \cite{attene2010lightweight}, \cite{attene2013polygon}, \cite{cherchi2020fast}, and \cite{zhou2016mesh}.

%
%
%

%% file: conclusion.tex
\section{Conclusion and Future Work}
\label{sec:conclusion}

In this paper,  a parallel feature-preserving mesh offsetting framework with variable distance is proposed. Our method can efficiently generate an offsetting mesh with smaller mesh size, and also can achieve high quality without gaps, holes, and self-intersections. The proposed method has been tested on the quadmesh dataset to show the robustness and efficiency.

There are several future research directions in this topic. For input meshes with  extremely poor quality requiring inward offsets, our method can still produce offset surfaces. However, in some regions, feature reconstruction should be improved. This challenge will be addressed by increasing the mesh density locally in the future. Furthermore, we also would like to to generate the offset mesh directly from spatial coverage without relying on TetWild.

\input{sample-bibliography.tex}

%% file: sample-bibliography.tex
\bibliographystyle{ACM-Reference-Format}
\bibliography{sample-bibliography}